\documentclass[12pt,a4paper,final]{iopart}

\expandafter\let\csname equation*\endcsname\relax
\expandafter\let\csname endequation*\endcsname\relax
\usepackage{iopams}
\usepackage[version=4]{mhchem}
\usepackage{upgreek}
\usepackage{subcaption}
\usepackage{graphicx}
\usepackage[breaklinks=true,colorlinks=true,linkcolor=blue,urlcolor=blue,citecolor=blue]{hyperref}
\usepackage{cite}
\usepackage{placeins}
\newcommand{\excludePart}[1]{}
\newcommand{\invasivenessPaper}[1]{#1}

\begin{document}

\title[Invasiveness of E-FISH]{Invasiveness of pico- and nanosecond E-FISH on plasma bullets in nitrogen}

\author{A.A.A. Limburg$^{1}$, T.E.W. Keur$^{1}$, R.F.E. Pleijers$^{1}$ and S. Nijdam$^{1}$}
\address{$^1$Department of Applied Physics, Eindhoven University of Technology, PO box 513, 5600 MB Eindhoven,
The Netherlands}
\ead{a.a.a.limburg@tue.nl}

\vspace{1cm} 

\excludePart{\interfootnotelinepenalty=10000
\freefootnote{Parts of this chapter are published as: \\ 
\noindent A.A.A. Limburg, T.E.W. Keur, R.F.E. Pleijers and S. Nijdam  ''Invasiveness of pico- and nanosecond E-FISH on plasma bullets in nitrogen.'', under review in Plasma Source Science and Technology, 2024}}

\invasivenessPaper{\begin{abstract}}
The electric field is the driving force behind every plasma. Electric field induced second harmonic generation (E-FISH) is a diagnostic able to obtain the electric field with high temporal and spatial resolution, is considered non-invasive and can be applied to almost any type of plasma with high sensitivity. However, the high power laser beam used as a probe in this technique, can interact with the gas and induce charges, which can subsequently influence the plasma. In this work, E-FISH is applied on non-thermal pulsed plasma jets in \ce{N2} flowing into atmospheric air. In these jets, ionization fronts propagate along the axis of the jet, which are highly reproducible and periodic. This allows for phase resolved measurements. A nanosecond and a picosecond pulsed laser, both operating at 1064\,nm, are used as sources. For the first time, the obtained E-FISH signals measured with both lasers are compared to each other. The results deviate significantly between the two lasers, which can be explained by laser induced guiding of the streamers. This is observed by taking ICCD images of the plasma trajectory. At the position where the plasma crosses the laser beam path, the plasma branches. This reveals that E-FISH is also invasive under some conditions. The profiles obtained with the picosecond laser are in good qualitative agreement with previous coherent Raman scattering-based four-wave mixing results on the same plasma source and therefore the picosecond laser is considered non-invasive. In future E-FISH measurements, the influence of the laser beam on the E-FISH signal should be taken into account to prevent changing the plasma behavior. By decreasing the laser power or using a shorter laser pulse, successful measurements can be performed. 
\invasivenessPaper{\end{abstract}}

\section[Introduction]{Introduction}
Laser diagnostics allow for localized, unobstructed and direct detection of all kinds of plasma parameters. For example, Thomson scattering, which is the scattering of laser light on free electrons, is used to determine electron density and electron temperature~\cite{Hubner_2015}. The electric field is the driving force behind every plasma. It determines the production of energetic particles, energy losses, gas temperature and chemical activity~\cite{Patnaik2017}. Therefore, information on the spatial and temporal distribution of the electric field could, for example, be used to enable gas temperature control throughout a plasma, which is essential in medical applications. A tool for measuring the electric field of discharges has been desired by the plasma community for decades. Several methods are used in previous studies. However, direct electric field measurements on transient discharges are often invasive or limited by photon emission and assumptions~\cite{Kozlov2007}. Electric field measurements based on the Stark effect are, for instance, based on emission spectroscopy. Therefore, the range over which the field can be determined is limited to the spatial and temporal range in which light is emitted by the plasma during a discharge~\cite{Sobota2016, Kozlov2001}. Another approach is to introduce a birefringent target surface~\cite{Tschiersch_2014,Slikboer2018}. The electric field of the plasma, impinging on the target, changes the refractive index of the target material. By measuring this change, the electric field of the plasma can be determined. These measurements are highly relevant in the material science industry or plasma medicine, in which a target is used. However, the electric field of the discharge is influenced by the presence of the target~\cite{Slikboer2016}. 

In this work, an optical probe measurement technique to measure electric fields that overcomes these disadvantages, is tested and developed. This technique, called electric-field induced second harmonic generation (E-FISH), has just become available, is easy to implement and shows great promise for a wide variety of plasmas~\cite{FirstEFISHplasma,EFISHVectormeasurements}. 
A nanosecond pulsed and a picosecond pulsed Nd:YAG laser, both emitting at 1064\,nm, are used as sources and non-linearly interact with an electric field. In this way, information on the spatial and temporal development of the field strength and polarization is obtained. The E-FISH technique is tested on plasma bullets in N$_2$ flowing into atmospheric pressure air.
Plasma bullets are guided streamer discharges. The high reproducibility of guided streamer discharges in pulsed plasma jets, which allows for phase resolved measurements, is generally ascribed to a memory effect, where
the seed electrons for the next discharge are provided by remnants of the previous
discharges~\cite{PlasmaJetPlume, Hofmann2012,10.1063/1.5031445}. The bullets cross the beam path in the focus. 

So far, laser diagnostics, including E-FISH, are generally considered non-invasive when applied to plasma~\cite{Goldberg_2022}. This makes them advantageous over most other plasma diagnostics. However, laser beams can induce the guiding of streamer discharges, such as plasma bullets, due to photoionisation or photodetachment of the medium through which the bullets move~\cite{LaserPositiveStreamerInteraction,LaserStreamerInteraction}.  If this occurs, the electric field profile of the plasma can change during the E-FISH measurement due to the laser beam. As a consequence, these measurements can no longer be deemed accurate and the technique has to be considered invasive. This would make E-FISH unsuitable to gain understanding about the discharge itself. In a recent paper of Nakamura~\cite{Nakamura2024} on E-FISH, the occurrence of laser induced streamers is described, but these discharges do not affect the E-FISH signal. 

In this work, the knowledge about the limitations and the possibilities of E-FISH as a new method to determine electric fields is extended by analyzing the impact of the laser beam on a plasma bullet. The effect of the pulse duration of this beam is investigated in regard to the accuracy of obtained electric field measurements, as well as the invasiveness on the plasma.  

In order for a non-linear interaction, such as second harmonic generation, to occur, a high intensity input beam is needed. Moreover, the E-FISH signal intensity scales as the square of the input intensity. 
Consequently, short, high energy laser pulses of the order of hundreds of picoseconds or less are preferred. Another advantage of shorter laser pulses is that the pulse energy, which is constrained by laser-induced breakdown~\cite{Demichelis1969}, is lower, while the beam intensity is higher. 
Laser-induced breakdown can influence the measured second harmonics (SH), either by disturbing the to-be-measured electric field or by generating light in the wavelength range for which the detector is sensitive. 
Furthermore, the time resolution of the E-FISH measurements is limited to the pulse duration of the laser source. Thus, for transient discharges, of which the electric field can vary on sub-nanosecond timescales, a nanosecond pulse duration may prove limiting. 
The E-FISH technique has been applied to electric fields in air using nanosecond, picosecond and femtosecond lasers~\cite{ Adamovich_2020,FirstEFISHplasma, EFISH_second}. However, the intrusiveness of the laser beams as a function of the pulse duration on the plasma was not experimentally assessed. 

This paper reveals that nanosecond E-FISH can be invasive when applied to plasma bullets and results need to be critically assessed. Furthermore, a picosecond laser beam can induce branching of the bullets. This will be a crucial extension to the fundamental knowledge of the E-FISH technique as it largely determines the accuracy of the results.

\excludePart{First, the experimental set-up is introduced in section two. In section three, electric field measurements on the plasma bullet and plasma images are presented and discussed. Finally, the results are summarized and conclusions are drawn in section 4.}

\section[Experimental setup]{Experimental setup} 

\begin{figure}
    \centering
    \includegraphics[width=1\textwidth]{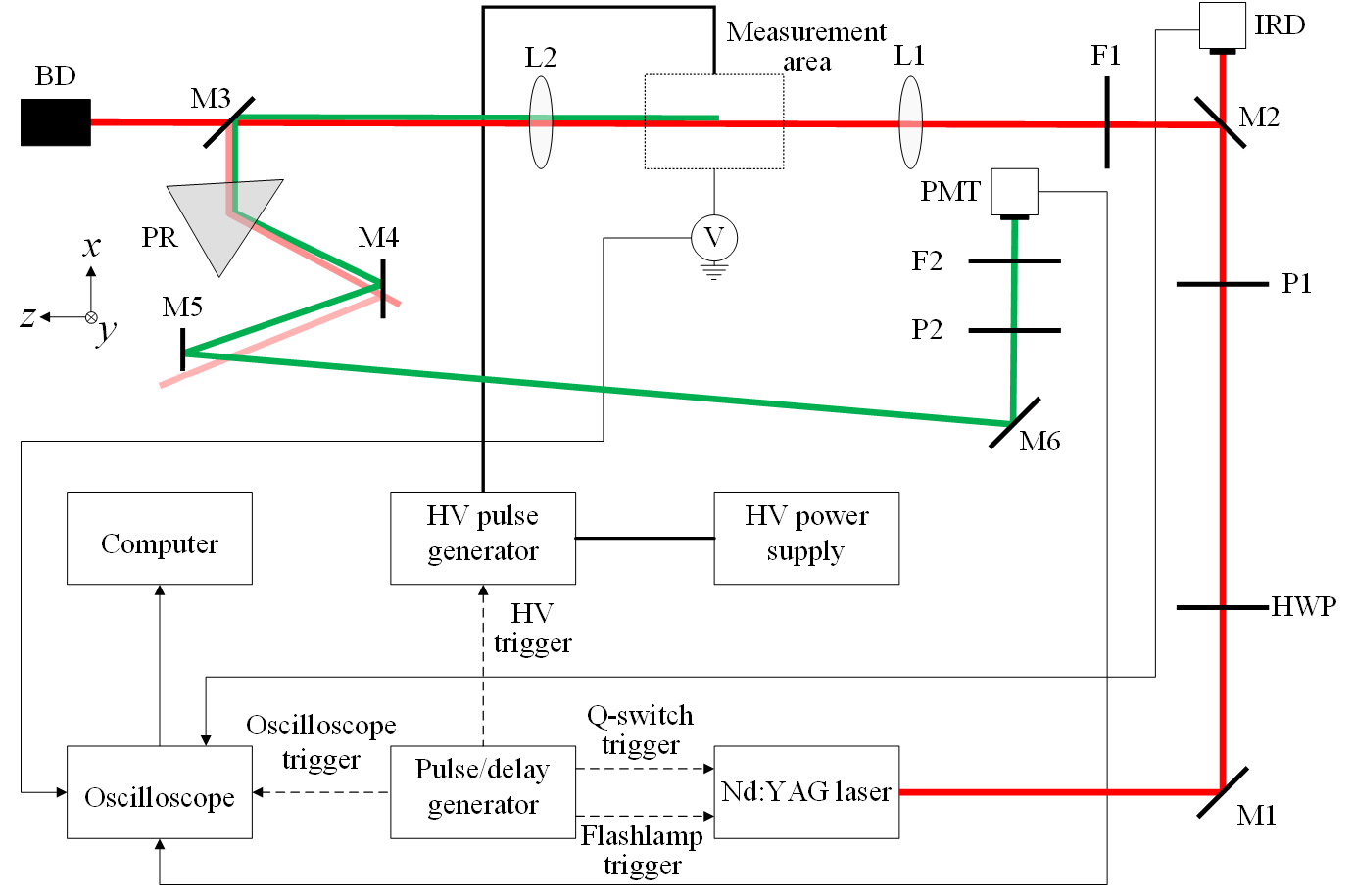}
    \caption{Schematic drawing of the experimental setup used for E-FISH measurements. The infrared laser beam is indicated in red, while the SH beam is indicated in green. M1: dielectric mirror; HWP: 1064\,nm half-wave plate; P1: 1064\,nm polarizer; M2: backside polished mirror; F1: 1064\,nm bandpass filter; L1 \& L2: plano-convex lenses; M3, M4, M5: dichroic mirrors (532\,nm reflecting, 1064\,nm transmitting); PR: prism; M6: dielectric mirror; P2: 532\,nm polarizer; F2: 532\,nm bandpass filter; BD: beam dump; IRD: infra-red detector; PMT: photomultiplier tube. The infra-red light is focused by L1 and collimated by L2. Furthermore, after the SH light is generated, both beams are colinear until they are separated by the prism.}
    \label{fig:ExpSetup_ExperimentalSetup}
\end{figure}

\begin{figure}
    \centering
    \includegraphics[width=\textwidth]{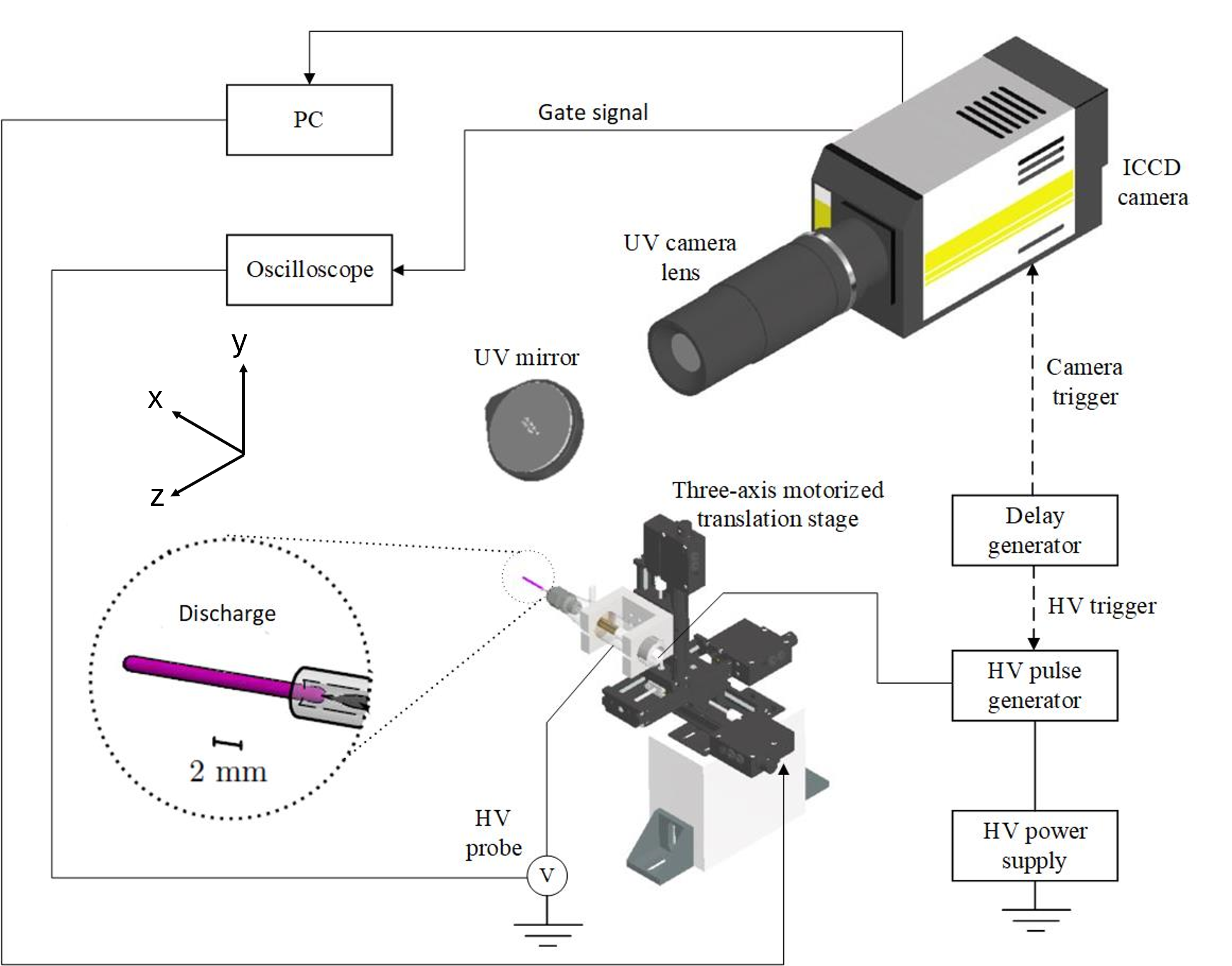}
    \caption[]{Schematic representation of the set-up used to take the ICCD camera images in Figure \ref{fig:ExpSetup_ExperimentalSetup}. The camera is triggered using the delay generator, and it in turn triggers the oscilloscope.}
    \label{fig:Cam_setup}
\end{figure}

\begin{figure}
    \centering
    \includegraphics[width=1\textwidth]{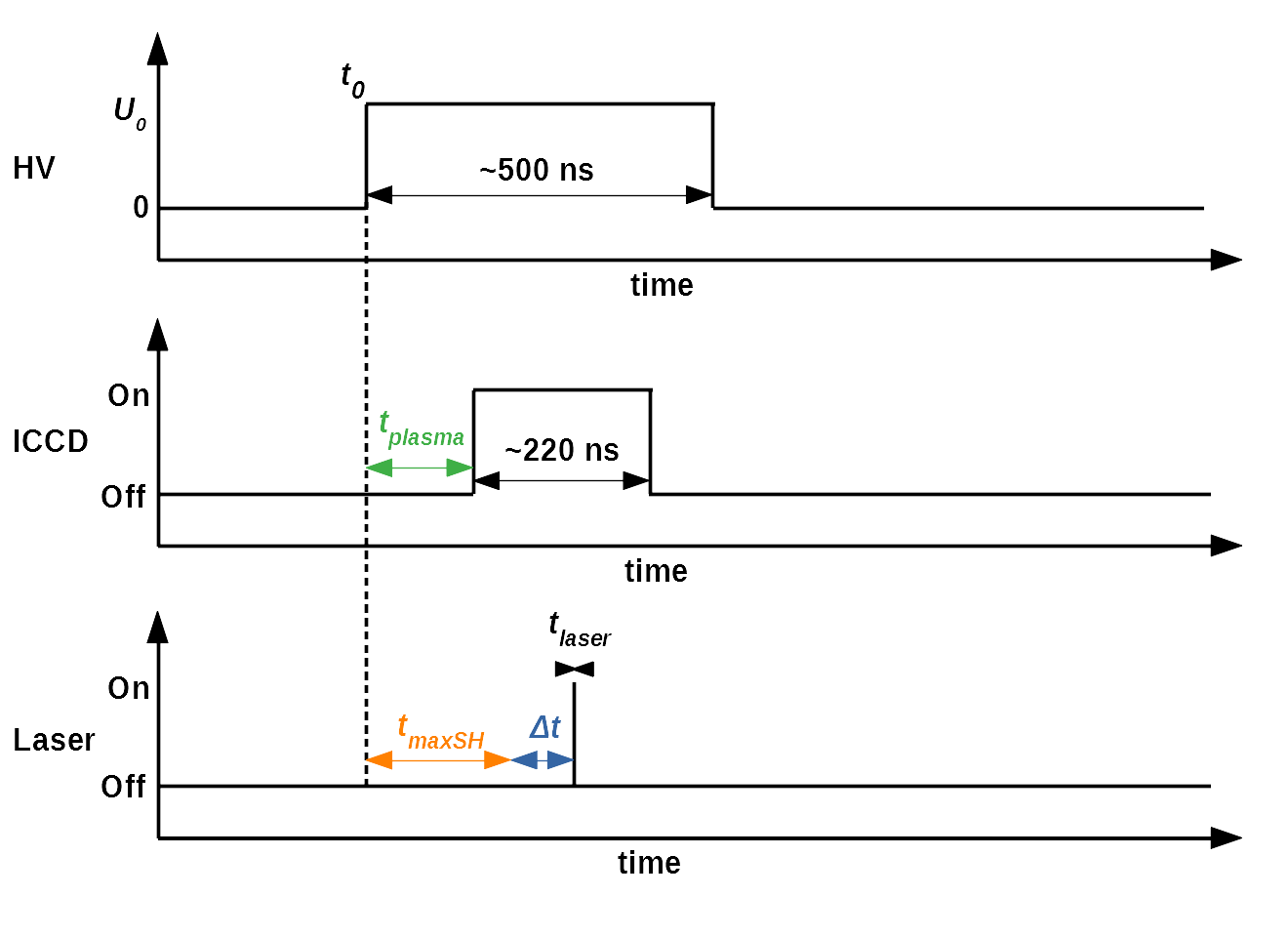}
    \caption[]{Schematic representation of the HV, laser and camera pulses as they are defined for this set-up. $t_0$ is defined as the start of the HV pulse. $t_{plasma}$ is the time at which the  plasma is first observed. $\Delta t$ is the variable laser pulse delay, which is centered around the time where the maximum SH intensity is measured $t_{maxSH}$. Timings are not to scale.}
    \label{fig:Timing_scheme}
\end{figure}

In figure~\ref{fig:ExpSetup_ExperimentalSetup}, a schematic of the used setup is presented. 
Two different Nd:YAG lasers are used in this study, both emitting at 1064\,nm: a Quantel Q-smart 450, which emits 6\,ns pulses at a frequency of 20\,Hz and an Ekspla 312 which emits 150\,ps pulses beam at a frequency of 10\,Hz. 
A laser beam emitted by one of these sources interacts non-linearly with an electric field, which results in frequency doubled light (SH). The input polarization of the laser light is controlled to achieve maximum second harmonic generation, e.g. the input light polarization is parallel to the to-be-measured field direction. Subsequently, the laser beam is focused into the measurement area using a plano-convex lens of 500\,mm focal length. Characteristics of the used plasma source can be found in~\cite{MarcVanDerSchansPhD}. 
To summarize: 
a pulsed plasma jet is generated by applying high voltage (HV) pulses to the plasma source using a HV power supply (Spellman SL50N150X4256), which is gated by a HV pulse generator (DEI PVX-4110), in pure N$_2$ (99.999\%), injected via a feed gas inlet at a volumetric flow rate of 1 standard litre per minute (slm). The HV pulses are set to be 500\,ns wide, with rise and fall times of about 40\,ns. 
The discharges are produced at a repetition frequency of 3\,kHz at a fixed voltage amplitude of 8\,kV. For precise positioning of the plasma bullets inside the laser beam, a motorized translation stage is used (Zaber Technologies T-XYZ-LSM050A). 
The shape of the plasma is monitored by an ICCD camera (Andor iStar CCD DH334T-18U-E3) with a Nikon UV 105\,mm f/4.5 lens, which is shown in figure~\ref{fig:Cam_setup}. The gate of the camera is either set to 2\,ns or 170\,ns/220\,ns, to capture the propagation of a single bullet and of the bullet path, respectively. The obtained ICCD images are post-processed using Matlab by subtracting the average background.
A gamma correction factor $\gamma_c$ is applied to increase the visibility of the plasma~\cite{BULL201499}. The nozzle and the colors were added in post-processing for illustration purposes. 
Since the laser and the electric field are both pulsed, it becomes necessary to align them in time, as indicated in figure~\ref{fig:Timing_scheme}. In order to split the second harmonic (SH) from the input laser beam, a combination of a prism and one dichroic mirror is used. The SH intensity is measured by a photomultiplier tube (Hamamatsu H6779-04) and an infra-red detector (Thorlabs DET36A/M) monitors the laser output intensity. The laser intensity in the focus of the beam is about $10^{9}$\,W/cm$^2$ and $10^{10}$\,W/cm$^2$ at maximum power for the nanosecond and picosecond laser, respectively.
All measurements are conducted in atmospheric pressure air at room temperature.

The distance between the needle electrode tip and the laser beam focus is
set to $4.0\pm 0.1$\,mm. \excludePart{Three}\invasivenessPaper{Two} different E-FISH measurements are performed:
\begin{enumerate}
    \item Radial scan: In order to obtain the radial field distribution, the $y$-position of the plasma source is changed from $-600$ till $+600\,\upmu$m in steps of 50\,$\upmu$m.
    \excludePart{\item Temporal scan: In order to determine how the electric field and the branching behavior vary over time, the laser pulse delay $\Delta t$ is varied from $-11$ till $+11$\,ns in steps of $1$\,ns. }
    \item Laser intensity scan: This scan is performed to investigate the effect of varying laser input intensity on the E-FISH measurements and the branching. According to \invasivenessPaper{\cite{MarcVanDerSchansPhD}, the 'effective' intensity of a Gaussian beam is given by
\begin{equation}
\label{eq:effective_intensity}I^\mathrm{GB} =\mathcal{P}/a_\mathrm{eff},
\end{equation} 
where $\mathcal{P}$ is the laser power and $a_\mathrm{eff}$ the effective area, defined as 
\begin{equation}\label{eq:App_EffectiveArea}
        a_\mathrm{eff} = \frac{\pi w_0^2}{4},
\end{equation}
with $w_0$ the beam waist.
Thus,}\excludePart{equations~\ref{eq:App_EffectiveArea} and \ref{eq:effective_intensity} in the appendix,} the laser intensity can be varied by either changing the laser power or the beam waist. The input power is varied by changing the angle orientation of the HWP as is shown in figure~\ref{fig:Power_fit} in the appendix. Furthermore, several focal length lenses are used to focus the beam, resulting in a different laser intensities around the focus.   
\end{enumerate}

\section[Results and discussion]{Results and discussion}

\subsection{The branching process}
\begin{figure}
    \centering
    \begin{subfigure}[b]{0.25\textwidth}
        \includegraphics[width=\textwidth]{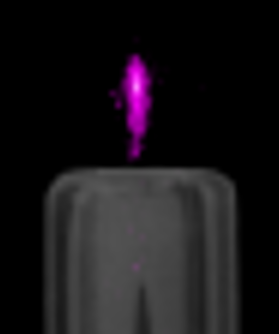}
        \caption{No laser}
        \label{fig:bullet_prop_a}
    \end{subfigure}
    \begin{subfigure}[b]{0.25\textwidth}
        \includegraphics[width=\textwidth]{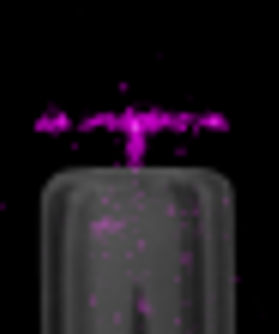}
        \caption{Nanosecond laser}
        \label{fig:bullet_prop_b}
    \end{subfigure}
    \begin{subfigure}[b]{0.25\textwidth}
        \includegraphics[width=\textwidth]{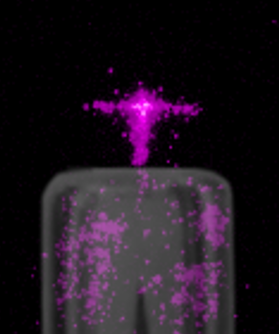}
        \caption{Picosecond laser}
        \label{fig:bullet_prop_c}
    \end{subfigure}
    \caption{ICCD images for 2\,ns gate time of the plasma bullet propagating without the presence of a laser beam (a), with the nanosecond laser (b), and with the picosecond laser (c) turned on. In these images, the laser travels from right to left, while the plasma moves from bottom to top. $\gamma_c = 1.6$ for (a) and (b), and 1.15 for (c).}
    \label{fig:bullet_prop}
\end{figure}

In figure~\ref{fig:bullet_prop_a}, the plasma bullet propagation is shown without a laser beam present. See figure~\ref{fig:Matrix_Loff}\invasivenessPaper{ in the appendix} for additional images. Inside the main channel, the remnants of previous discharges guide the new streamer, leading to the straight undisturbed plasma bullet trajectory. This is known as the ”memory effect”~\cite{MarcVanDerSchansPhD,Nijdam_2014,Li_2018}. 

In figures~\ref{fig:bullet_prop_b} and~\ref{fig:bullet_prop_c}, ICCD images obtained with the same settings are shown with the nanosecond and picosecond laser turned on, respectively. See figures~\ref{fig:Matrix_Jetscan_500mm_ns} and~\ref{fig:Matrix_Jetscan_500mm}\invasivenessPaper{ in the appendix} for additional images. It can be observed that the path of the jet is changed; the plasma is guided into the laser beam path. Furthermore, if branching into the laser path occurs, it always occurs in both directions. 
Moreover, we identified that the polarization direction of the input light does not matter for the branching. 
Also, the branching process is very stochastic as can be seen in figure~\ref{fig:Matrix_Jetscan_stoch} in the appendix.

\begin{figure}
    \centering
    \includegraphics[width=0.7\textwidth]{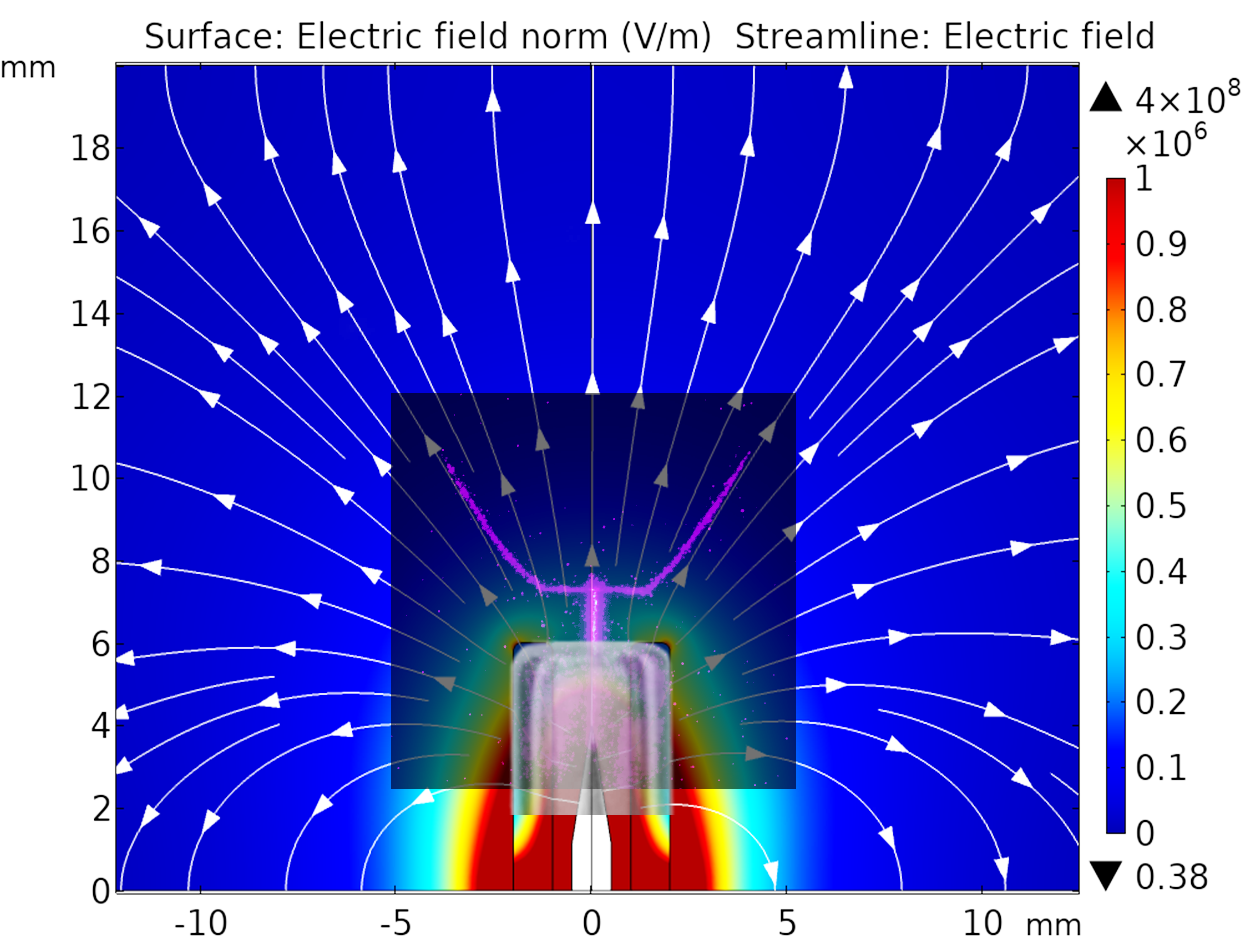}
    \caption[]{COMSOL Multiphysics model of the electric field streamlines overlaid with an ICCD image with gate time 220\,ns of the bullet trajectory for the picosecond laser.}
    \label{fig:ps_Efield_stream}
\end{figure}

From figure~\ref{fig:ps_Efield_stream}, showing the full trajectory of the branching bullets for the picosecond laser, we can observe that the bullets branch into the laser beam path and at a certain point escape from this path. The bullet follows the electric field lines before and after the branching into the laser beam path. The same holds when using the nanosecond laser as can be seen in figure~\ref{fig:streamlines_ns}\invasivenessPaper{ in the appendix}. Moreover, we observed that branches leave the picosecond laser beam closer to the plasma propagation axis compared to using a nanosecond laser. The minimum gate width of the camera is 2\,ns, which makes it difficult to assess how long it takes before the branches are created. The branches are at least developed within 2\,ns. Given the average speed of the bullets, 10$^5$\,m/s~\cite{MarcVanDerSchansPhD}, the bullet will move approximately 600\,$\upmu$m during a nanosecond laser pulse and 15\,$\upmu$m during a picosecond laser pulse. Especially when using the nanosecond laser, the branches are able to develop during the measurement. However, the laser pulse has already past far before the bullet leaves the laser path, which takes about 20\,ns for the nanosecond laser and 14\,ns for the picosecond laser after branching. Therefore, the difference in guiding is likely due to the electron density induced by the laser diminishing over time, eventually becoming insufficient to confine the bullets, which occurs sooner for the picosecond laser because of the shorter pulse duration.

\begin{figure}
    \centering
    \includegraphics[width=0.5\textwidth]{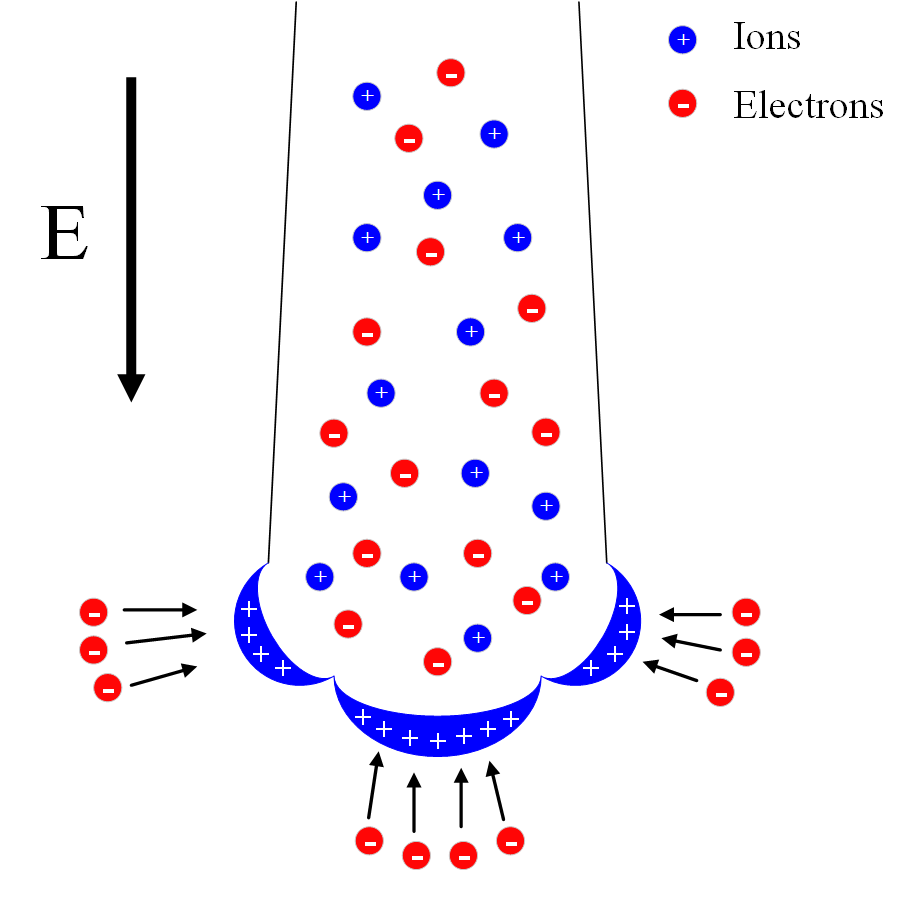}
    \caption{A schematic illustration of a branching event of a streamer discharge, due to the presence of a significant amount of electrons in the laser beam path.}
    \label{fig:Theory_Streamer_Branching}
\end{figure}

According to previous research, in gas mixtures with little photo-ionization a streamer discharge tends to travel into the direction of higher electron density~\cite{Nijdam_2014}. The branching process is shown in figure~\ref{fig:Theory_Streamer_Branching}.
The electron density in a regular streamer path is around $10^{14}$\,cm$^{-3}$~\cite{Nijdam2020}. We found that the plasma bullet velocity is lower when it branches, which can be explained by the fact that current must be distributed over the three paths at the intersection with the laser. 

From research on the memory effect, it is determined that electron densities of $10^8 - 10^9$\,cm$^{-3}$, which are far below the breakdown threshold, are already sufficient to guide the plasma in a direction nearly perpendicular to the background electric field direction~\cite{LaserPositiveStreamerInteraction}. In order to find the origin of the electrons present in the laser beam path, we need to discuss reactions that generate electrons in our setup.
 
Positive streamers in air propagate through photoionization of oxygen molecules by ultraviolet (UV) photons emitted by excited nitrogen molecules, because this provides the required free electrons in front of the streamer head. This works as follows: when a high energy electron moving in air collides with a nitrogen molecule and excites it, the nitrogen molecule can subsequently emit a photon with a wavelength of $98-102.5\,$nm. This photon in turn can ionize an oxygen molecule somewhere else, creating an additional electron~\cite{Wormeester_2010}:
\begin{equation}
\label{eq:photoionization}
\begin{split}
 &\ce{N2}^* \rightarrow \ce{N2} + h\nu \\
&\ce{O2} + h\nu \rightarrow \ce{O2}^+ + \text{e},   
\end{split}
\end{equation}
where $h\nu$ represents the photon energy. 
The process of photoionization stabilizes the discharge front in both the inception cloud phase and in the propagating streamer phase. Besides being created by photoionization, free electrons can also be left over from previous discharges, and can be created by electron detachment from negative ions. Also, external radiation like cosmic rays or bremsstrahlung photons from very fast runaway electrons can act as an electron source~\cite{Nijdam2020}.

 An electron can be released from a negative ion by irradiating the ion with a photon with an energy higher than the detachment energy. The photodetachment of oxygen ions is described as
\begin{equation}
\begin{split}
\label{eq:photodetachment}
&\ce{O3^-} + h\nu \rightarrow \ce{O3} + \ce{e^-}, \text{with $h\nu > 2.10$\,eV},\\
&\ce{O2^-} + h\nu \rightarrow \ce{O2} + \ce{e^-}, \text{with $h\nu > 0.45$\,eV},\\
&\ce{O^-} + h\nu \rightarrow \ce{O} + \ce{e^-}, \text{with $h\nu > 1.46$\,eV}.
\end{split}
\end{equation}
The photon energy of the used laser (1064\,nm) is 1.17\,eV. Therefore, only \ce{O2^-} is vulnerable to laser induced photodetachment in our setup, because of the higher electron binding energies of \ce{O^-} and \ce{O3^-} (above photon energy)~\cite{Hasani_2021}. For the laser power used in this research, $\ce{O2}^-$ is almost 100\% detached within the laser beam, taking into account the photodetachment cross-section~\cite{PhysRev.112.171}. Thus, photodetachment is the most likely source of the electrons and therefore the branching. 

\begin{figure}
    \centering
    \includegraphics[width=0.7\textwidth]{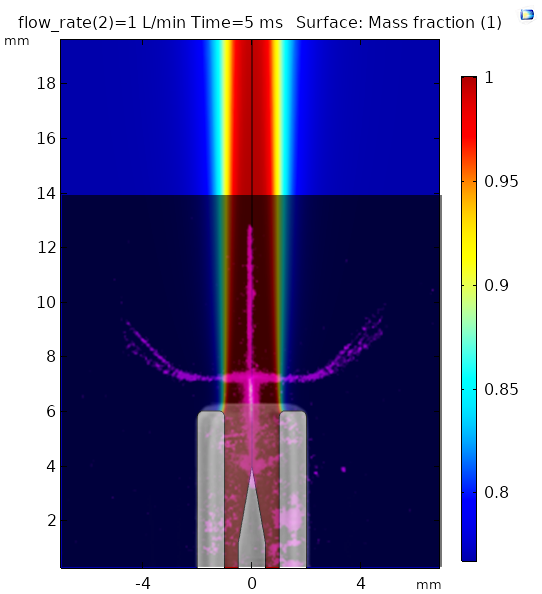}
    \caption[]{COMSOL Multiphysics model of the mass fraction \ce{N2} for the used plasma nozzle, overlaid with an ICCD image with gate time 170\,ns of a branching discharge for the nanosecond laser.}
    \label{fig:N2_branch_comp}
\end{figure}

Further away from the center of the studied plasma jet, the mass fraction of nitrogen decreases, as can be observed in figure~\ref{fig:N2_branch_comp}, and subsequently the oxygen concentration increases and more and more photoionization electrons become available, making the streamer less sensitive for variations in background electron density and therefore more difficult to laser guide~\cite{LaserPositiveStreamerInteraction,Nijdam2020}.
This might be the reason for the branches leaving the laser path, because in pure nitrogen, a low concentration of laser-detached electrons is already enough to guide the streamer. 
This effect is enhanced by the faster attachment of (laser-detached) free electrons at higher oxygen concentrations. Therefore, in high purity nitrogen, there will be more free electrons before the streamer arrives, which can also better guide the streamer.
Furthermore, the brightest branching occurs in the region where the mass fraction of \ce{N2} is the highest.

Finally, in theory the laser could also create free electrons in its beam path by direct ionization of gas molecules~\cite{LaserStreamerInteraction,LaserPositiveStreamerInteraction,10.1063/1.4769261,Zvorykin_2011}. 
The photon energy of the infra-red (IR) input laser is too low to overcome the ionization potential in a single-photon process, but multiple photons together do exceed this barrier and are able to ionize when absorbed simultaneously. 
For photoionization by the input laser, 11 photons are needed to ionize an oxygen molecule and 14 photons for nitrogen, which are the molecules that are mostly present. The needed laser intensity to cause breakdown due to photoionization using an IR laser is $10^{11}$\,W/cm$^2$~\cite{10.1117/12.2539007}. However in this research, the laser intensity in the focus of the beam is over one order of magnitude below this value, especially because we attenuate the beam.

To summarize: the guiding by the laser is most likely a combination of the ions present in the laser beam path due to the previous discharge (memory effect) and the laser detaching the electrons from these negative oxygen ions. 
The oxygen concentration is high enough to have these negative ions present and the laser intensity is probably too low to ionize oxygen and nitrogen molecules, but high enough to detach electrons from negative oxygen ions. 

\subsection{Impact of branching on E-FISH measurements}

\begin{figure}
    \centering
    \begin{subfigure}{\textwidth}
      \makebox[\textwidth][c]{\includegraphics[width=1.3\textwidth]{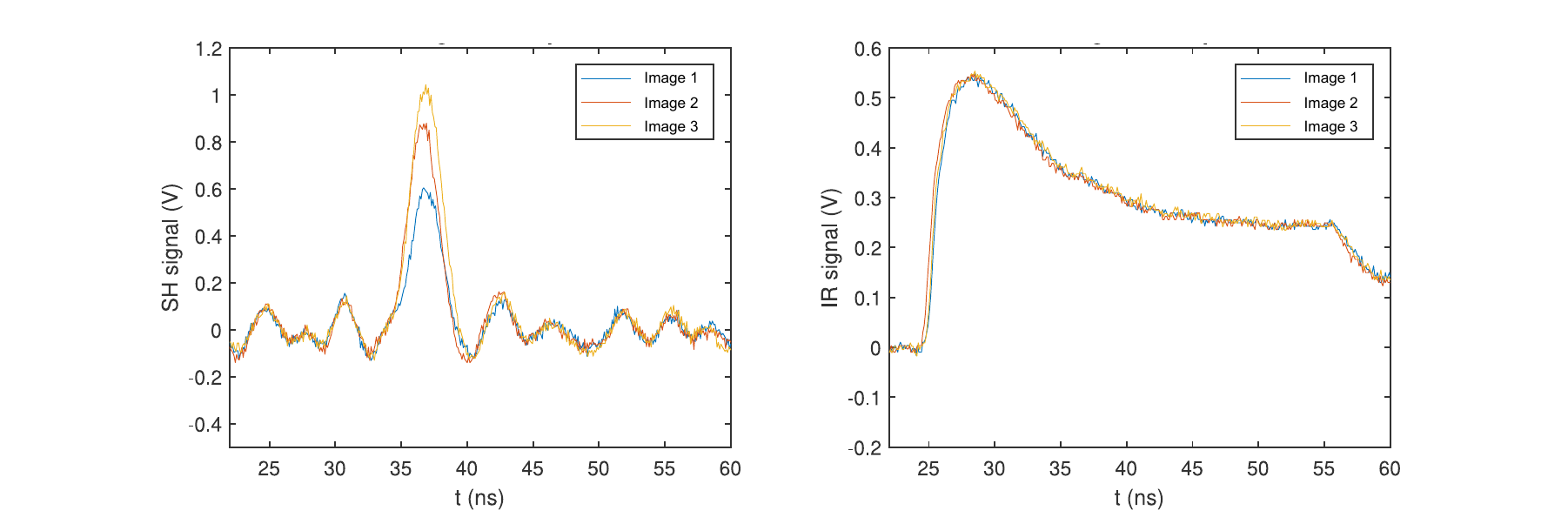}}
      \caption{}
        \label{fig:SH_peaks_steps}
    \end{subfigure}
    
    \begin{subfigure}{.8\textwidth}
      \includegraphics[width=\linewidth]{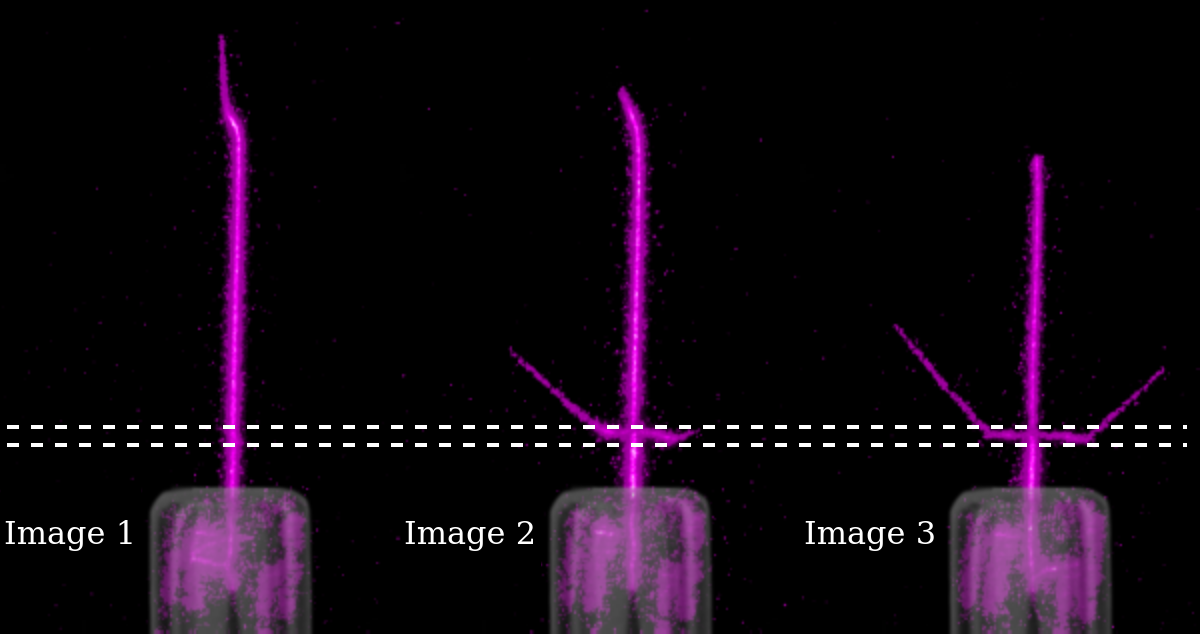}
      \caption{}
        \label{fig:hscan_branch_steps}
    \end{subfigure}
\caption{(a) The SH and IR signal corresponding to the radial electric field component using the picosecond laser, comparing three images with the same experimental parameters. (b) The corresponding ICCD images for 220\,ns gate time. In these images, the laser
travels from right to left between the two dashed lines, while the plasma moves from bottom to top. $\gamma_c = 1.15$.}
\label{fig:SH_branch_steps}
\end{figure}

In figure~\ref{fig:SH_branch_steps}, it can be observed that  the branching of the jet significantly affects the measured SH signal when using the picosecond laser. When branching occurs, an increase of about 80\% in maximum SH signal is measured for a constant IR signal and the same experimental conditions, where the SH signal is shown to decrease for less intense branching. This is the consequence of an increase in electric field length in the laser beam direction when the plasma bullet is traveling into the laser beam, and this field length being far below the coherence length. Similar behavior is observed when using the nanosecond laser. 
The branches at the end of the jet tend to the left, which is also the position of the post to mount the mirror in the imaging setup in figure~\ref{fig:Cam_setup}. This is likely the nearest electrical ground for the plasma. 

\subsubsection{Radial scan}
\begin{figure}
    \centering
    \begin{subfigure}{0.49\textwidth}
      \centering
    \includegraphics[width=\textwidth, trim=60 220 80 200, clip]{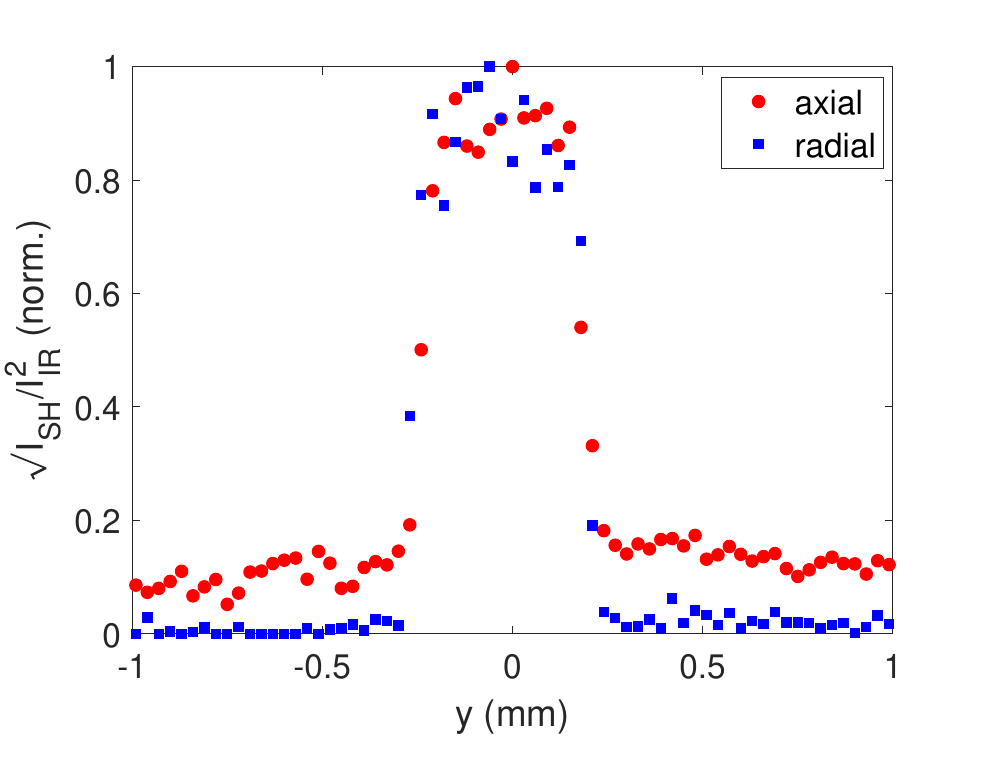} 
    \caption{}
     \label{fig:top_hat_a}
    \end{subfigure}
     \begin{subfigure}{0.49\textwidth}
      \centering
    \includegraphics[width=\textwidth, trim=60 205 80 200, clip]{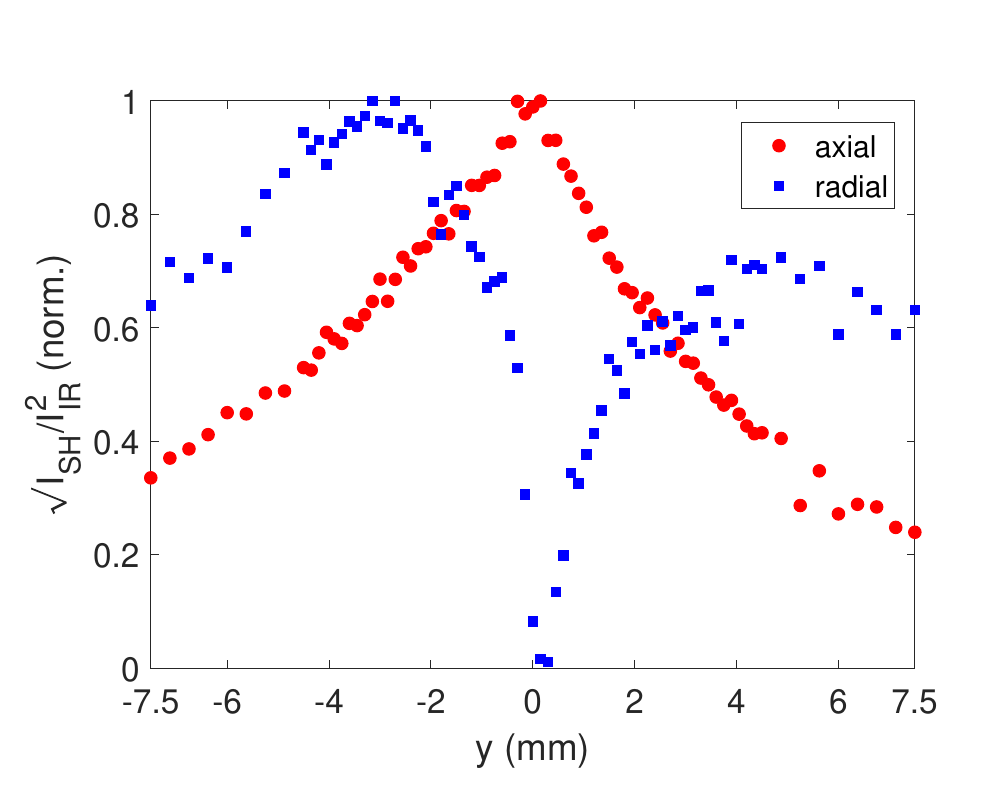} 
    \caption{}
     \label{fig:top_hat_b}
    \end{subfigure}
    \begin{subfigure}{0.8\textwidth}
      \includegraphics[width=\linewidth]{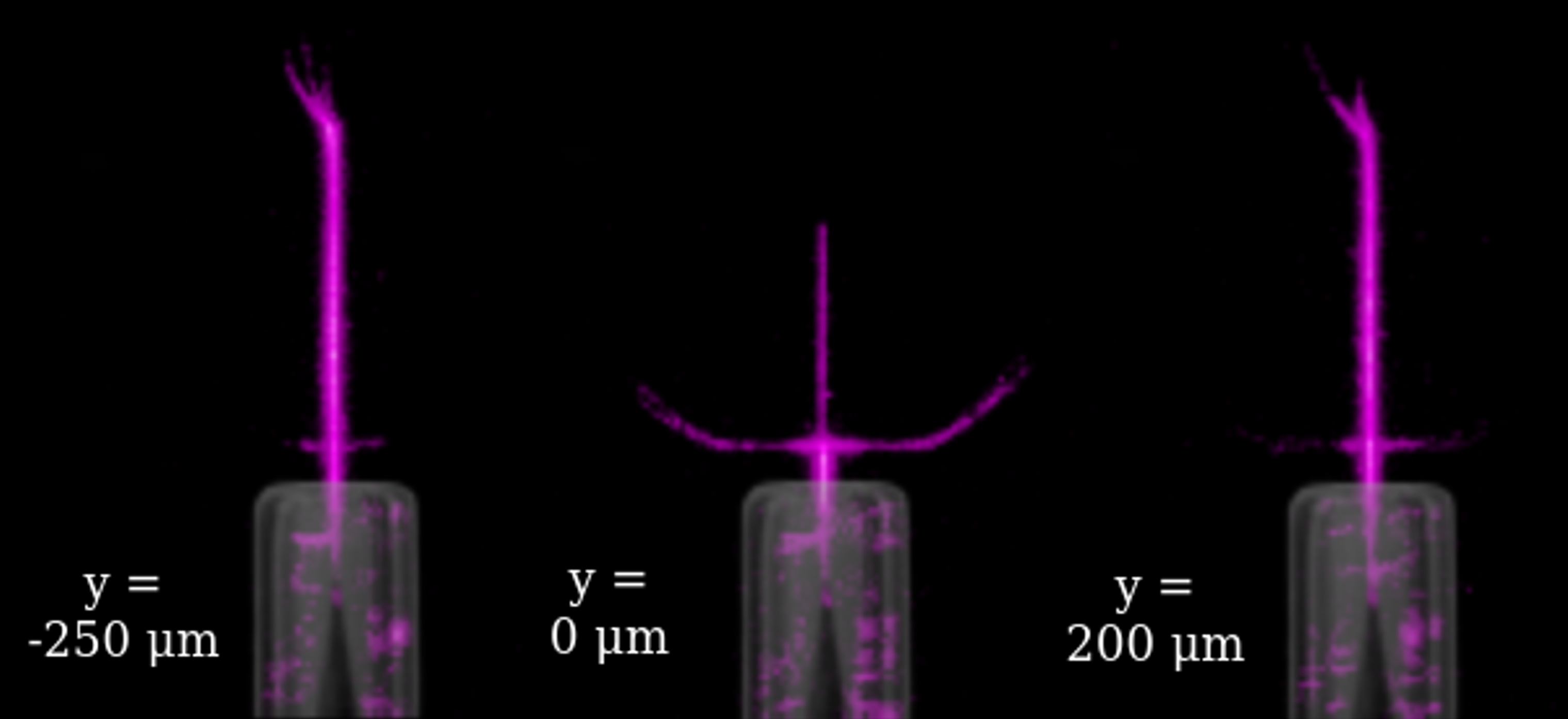}
      \caption{}
      \label{fig:top_hat_c}
    \end{subfigure}
\caption{(a) Radial profiles of the E-FISH signal for the axial and radial electric field, obtained with the nanosecond and picosecond laser. The electric field profile corresponding to the nanosecond laser shows an unexpected top hat profile. (b) The corresponding ICCD images for 170\,ns gate time for the nanosecond laser. In these images, the laser travels from right to left, while the plasma moves from bottom to top. $\gamma_c = 1.30$.}
\label{fig:Top_hat}
\end{figure}

Figure~\ref{fig:top_hat_c} shows that the branching is most apparent around $y = 0$\,$\upmu$m and decreases with increasing and decreasing $y$. Additional images can be found in figure~\ref{fig:Matrix_hscan_horpol_500mm_ns} in the appendix. The same holds for the ICCD images using the picosecond laser in figure~\ref{fig:Matrix_hscan_horpol_500mm}\invasivenessPaper{ in the appendix}. The SH signal measured for the axial and radial field components increases significantly when branching occurs, resulting in a ‘top hat’ profile when using the nanosecond laser, as shown in figure~\ref{fig:top_hat_a}. The profile is expected to gradually decrease for increasing $y$, but it seems constant for a range of about 500\,$\upmu$m. 
Branching was observed in a range of 450$\pm$50\,$\upmu$m. This exceeds the optical bullet diameter of $280 \pm 20$\,$\upmu$m~\cite{MarcVanDerSchansPhD}, but corresponds to the width of the top hat profile. The picosecond E-FISH measurements do not show the top hat profile, but the axial profile is slightly flattened around $y=0$. The difference is likely caused by a lower laser power, which still provides sufficient E-FISH signal. Additionally, the minimal movement of the bullet during the picosecond laser pulse could also play a role.

\excludePart{\subsubsection{Temporal scan}
\label{subsubsection bullets temporal scan}
The ICCD images at different laser delays in figures~\ref{fig:Matrix_time_pow1_3} and~\ref{fig:Matrix_tscan_horpol_500mm} show branching from $\Delta t$ is $-9\,\text{ns to } 7\,\text{ns}$ for the nanosecond laser and from $\Delta t$ is $-3\,\text{ns to } 11\,\text{ns}$ for the picosecond laser. 
E-FISH measurements were performed for three different values of $y$: 0, 233 and 465\,$\upmu$m and a time range of 40\,ns for the nanosecond laser and 100\,ns for the picosecond laser, where $\Delta t$=0 corresponds to the position of the peak of the axial profile.

First, the nanosecond laser is used to obtain the temporal E-FISH profiles corresponding for the axial and radial electric field components of the plasma bullet. In figures \ref{fig: horizontal time scan nanosecond} and \ref{fig: vertical time scan nanosecond}, the E-FISH signal for the axial and radial field components are shown as a function of time. For the two non-zero values of $y$, the radial profile has a small tail. This tail is expected when looking at figure \ref{fig:Theory_Streamer_Branching} and earlier obtained results on the same plasma source in~\cite{MarcVanDerSchansPhD}. Before the bullet head reaches the beam path, clearly no radial field is present, which is again in agreement with previous research. The axial profile starts to build up earlier and also decreases earlier. The results for the radial and axial field look very similar, which is unexpected and probably due to branching. 

\begin{figure}
\centering
 \includegraphics[width=0.83\textwidth]{Figures/E-FISH_invasiveness/hor_time_scan_plot.png}
   \caption{Temporal profile of E-FISH signal for the axial electric field component at three different positions of $y$ obtained with the nanosecond laser.}
  \label{fig: horizontal time scan nanosecond} 
\end{figure}
\hfill
\begin{figure}
\hspace{1cm}
   \includegraphics[width=0.8\textwidth]{Figures/E-FISH_invasiveness/ver_time_scan_plot.png}
   \caption{Temporal profile of E-FISH signal for the radial electric field component at three different positions of $y$ obtained with the nanosecond laser.}
   \label{fig: vertical time scan nanosecond}
\end{figure}

The picosecond E-FISH results are only obtained on the jet axis and are shown in figure~\ref{fig: comparison axial profile}. The nanosecond and the picosecond results show similar peaks at similar time instances and also the behavior in front of the bullet is equal. Moreover, the tail of the bullet is visible, which is not the case in the nanosecond results. Both the profile for the axial and the radial field component show great qualitative similarities with previous results on the same plasma jet~\cite{MarcVanDerSchansPhD}.

\begin{figure}
\centering
\includegraphics[width=0.8\textwidth]{Figures/E-FISH_invasiveness/hor_vs_vert_time_scan_plot.png}
\caption{Temporal profile of E-FISH signal for the radial and axial electric field component obtained with the picosecond laser.}
\label{fig: comparison axial profile}
\end{figure}}

\subsubsection{Laser intensity scan}
\begin{figure}
\centering
\includegraphics[width=0.7\textwidth]{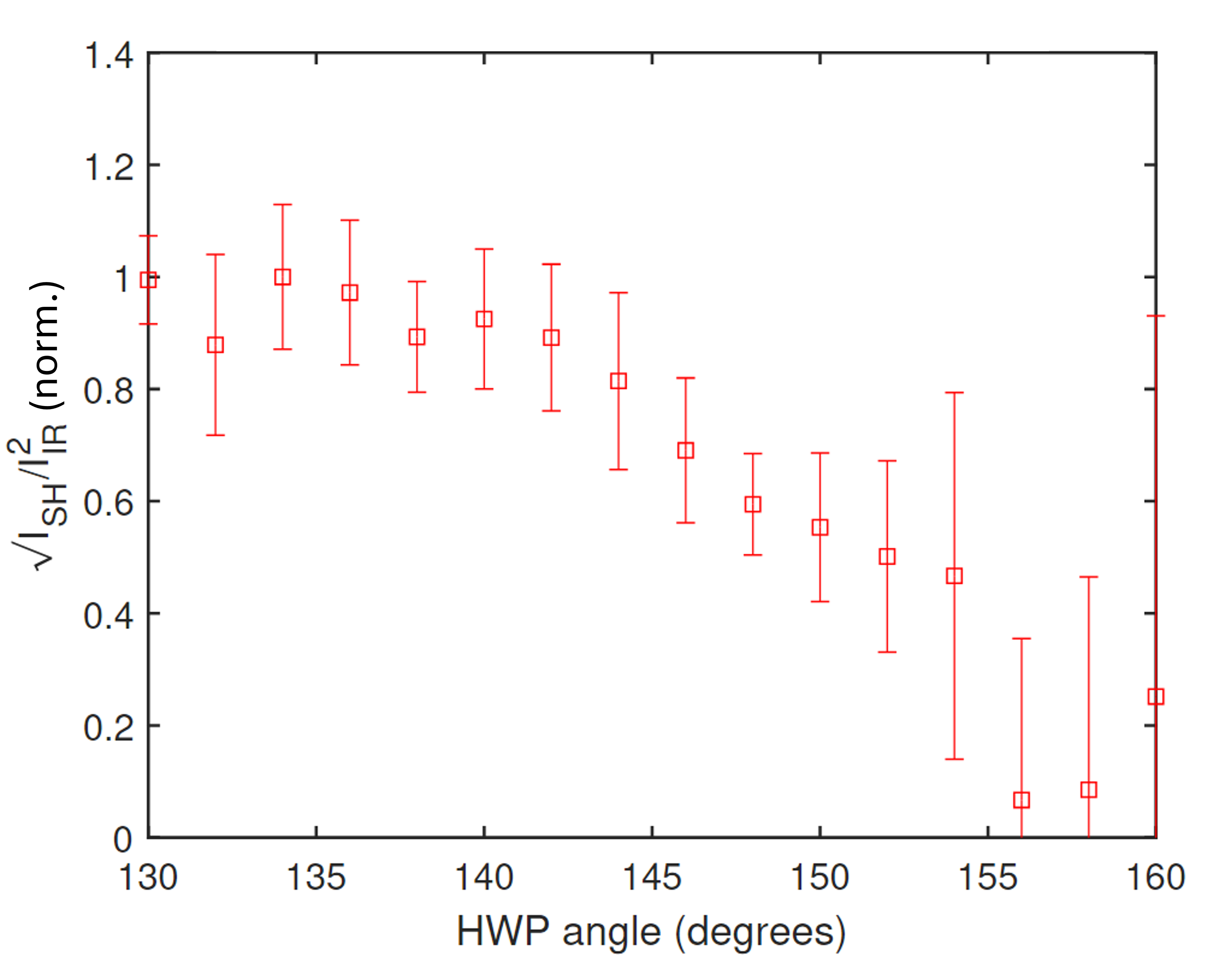}
\caption{Normalized E-FISH signal for varying HWP angles for the picosecond laser. These angles correspond to varying input laser powers according to figure~\ref{fig:Power_fit}\invasivenessPaper{ in the appendix}. The axial electric field component is measured.}
\label{fig: laser power scan}
\end{figure}

The laser power scan, shown in figure~\ref{fig: laser power scan}, reveals a decrease in the measured E-FISH signal for decreasing laser power, in contrast to the constant value expected from theory. The power where the decrease sets in corresponds to the point where the branching starts to decrease in intensity, as can be seen in figure~\ref{fig:Matrix_powscan_500mm}\invasivenessPaper{ in the appendix}. At the input power for which branching no longer occurs, the measurement error is too large to obtain a significant SH signal.

\begin{figure}
\centering
\includegraphics[width=0.8\textwidth]{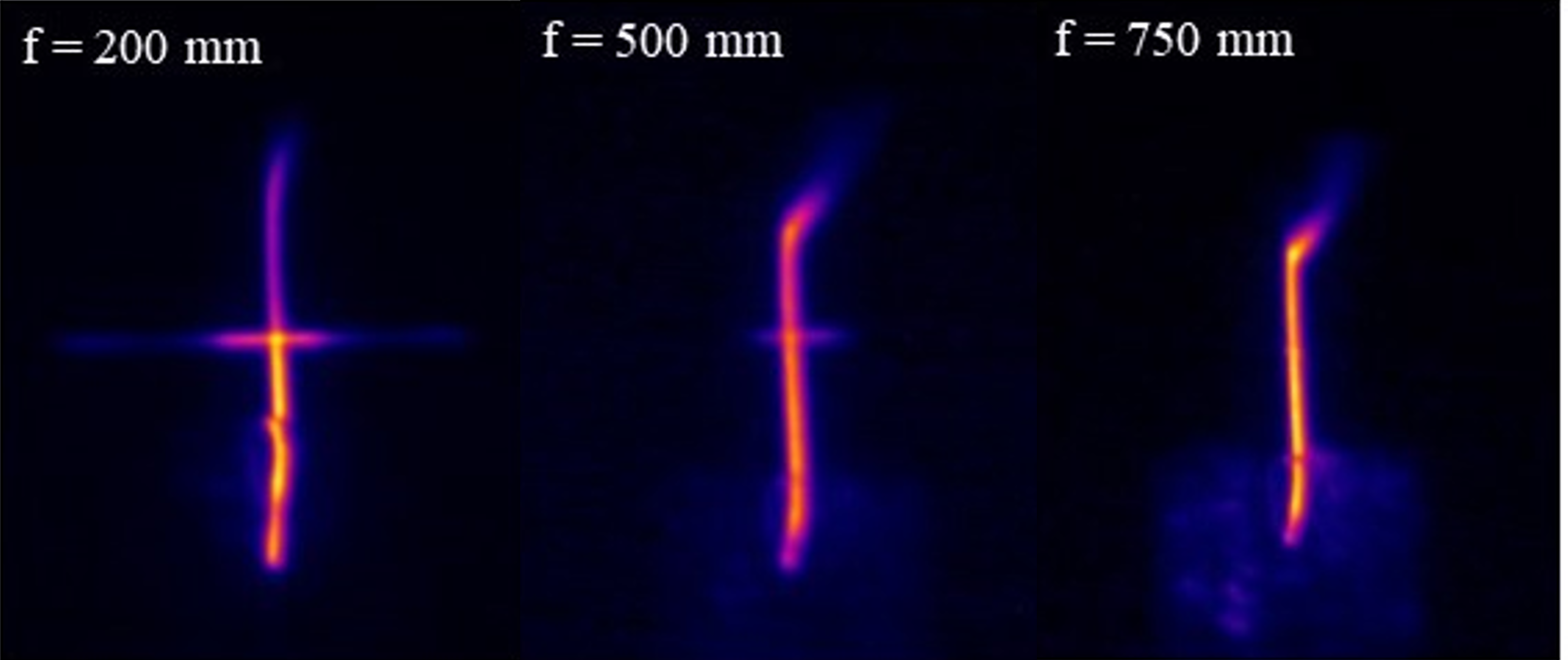}
\caption{ICCD images (obtained with a Stanford 4 Quick E instead of the Andor iStar camera) of the interaction between the plasma jet and the nanosecond laser for varying focal lengths $f$ of the focusing lens. The gate time of the ICCD was set at 220\,ns and all images are an accumulation
of 200\,frames. The laser travels from right to left, while the plasma moves from bottom to top.
Note that the nozzle and the needle electrode are not displayed in these images.}
\label{fig: focal length comparison}
\end{figure}

Furthermore, by varying the focal length of the focusing lens, shown in figure~\ref{fig: focal length comparison}, it can be observed that the amount of branching of the plasma jet increases for a more focused laser beam.

In order to show the impact of using a lower laser power, the alignment was improved by including more dichroic mirrors as indicated in figure~\ref{fig:ExpSetup_ExperimentalSetup}, such that a lower input laser power could be used while obtaining similar SH intensities. Subsequently, radial scans are obtained for the different focal length lenses using the nanosecond laser. In figures~\ref{fig:radialscan_focal1} and~\ref{fig:radialscan_focal2} the profiles are shown. The top hat profile can again be observed for the 200\,mm focal length lens, which is expected from the corresponding branching behavior in figure~\ref{fig: focal length comparison}. Fortunately, the 500\,mm focal length lens results look more similar to the expected profiles and the 750\,mm results are almost in agreement with literature~\cite{MarcVanDerSchansPhD}. There are two main differences: The first difference is the asymmetry in the measured E-FISH signal corresponding to the radial field component that is induced by interference effects, as explained in \invasivenessPaper{\cite{guo2024}}\excludePart{section~\ref{sec: interference}}. The second difference is that the value of the measured E-FISH signal for the radial field component is not zero at $y=0$, because of the limited spatial resolution when using a lens with a $750$\,mm focal length.

\begin{figure}
    \centering
    \begin{subfigure}[b]{0.75\textwidth}
        \includegraphics[width=\textwidth, trim=20 30 10 10, clip]{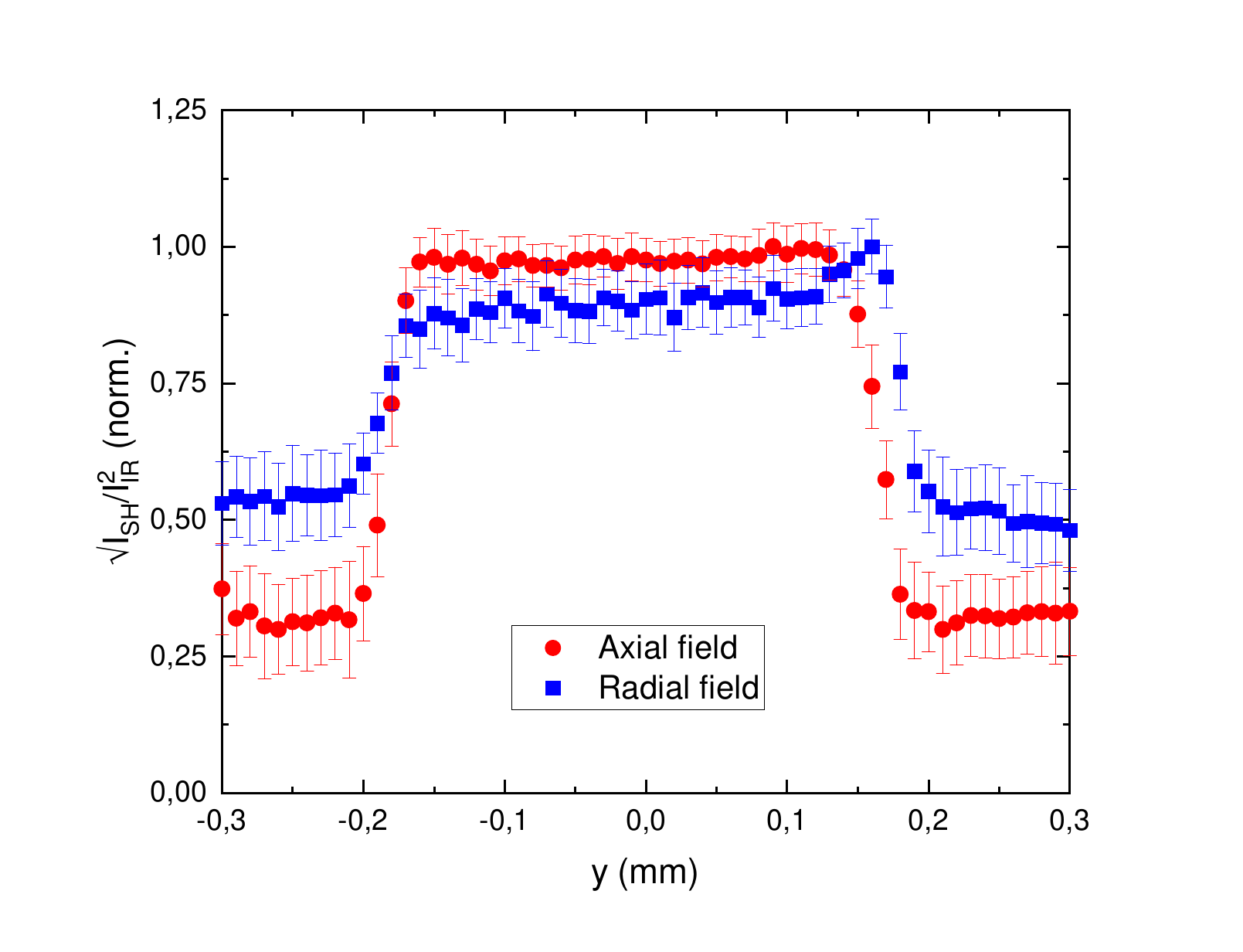}
        \caption{$f=200$\,mm}
        \label{fig:200mm}
    \end{subfigure}
     \begin{subfigure}[b]{0.75\textwidth}
        \includegraphics[width=\textwidth, trim=20 30 10 10, clip]{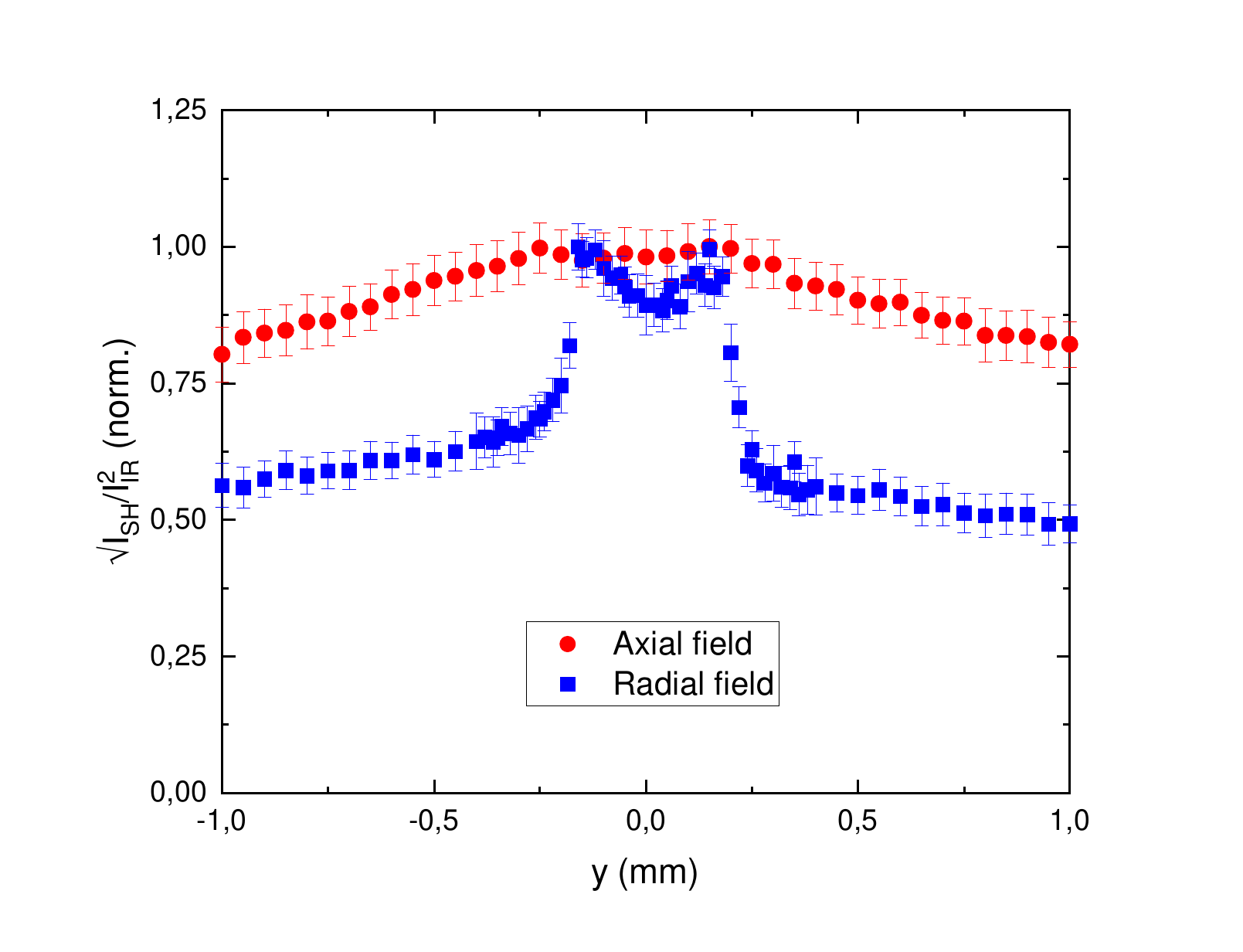}
        \caption{$f=500$\,mm}
        \label{fig:500mm}
    \end{subfigure}
    \caption{Normalized radial profiles of the E-FISH signal corresponding to the axial and radial electric field obtained using the nanosecond laser for varying focal lengths of the focusing lens.}
    \label{fig:radialscan_focal1}
\end{figure}

\begin{figure}
    \centering   
    \begin{subfigure}[b]{0.75\textwidth}
        \includegraphics[width=\textwidth, trim=20 30 10 10, clip]{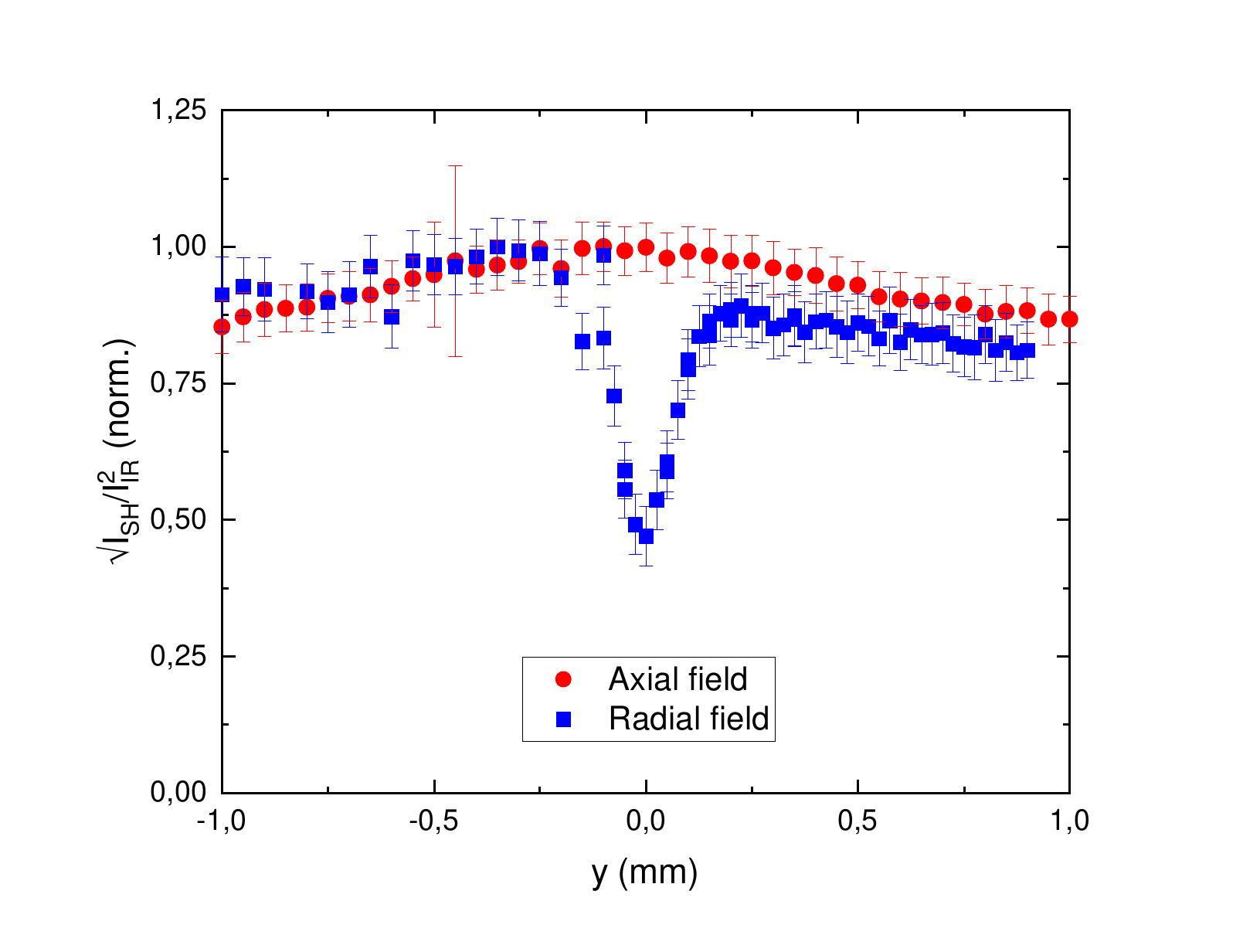}
        \caption{$f=750$\,mm}
        \label{fig:750mm}
    \end{subfigure}
    \caption{Continuation of normalized radial profiles of the E-FISH signal corresponding to the axial and radial electric field obtained using the nanosecond laser for varying focal lengths of the focusing lens.}
    \label{fig:radialscan_focal2}
\end{figure}

\section[Conclusion]{Conclusion}
This research reveals an important complication that can occur when performing E-FISH measurements on plasmas that move into the direction of the highest electron density. This issue is also likely present, albeit to a lesser extent, in static plasmas. We found that the probing laser beam, both for picosecond and nanosecond lasers, used during E-FISH measurements has a significant influence on the trajectory of an \ce{N2} ns pulsed plasma jet in atmospheric pressure air. This is due to branching of the plasma into the laser beam path, which is observed by imaging the plasma bullet trajectory. 
The electrons that induce the branching must be created through photodetachment reactions as the laser power is deemed too low for laser induced breakdown. After leaving the laser path, which is due to electrons shielding the streamer head as a result of a higher oxygen concentration, the streamer branches again follow the field lines of the background electric field.

The effect of pulse duration of the probing laser beam is investigated by comparing the branching behavior and the E-FISH signal obtained with a picosecond and a nanosecond laser. We determined that the branch formation takes place on a timescale below 2\,ns. Therefore, the formation occurs completely during a nanosecond laser pulse. It is shown that branches result in a higher SH intensity. For nanosecond E-FISH measurements, when the SH signal is measured at different radial positions in the plasma, the obtained profile is completely determined by the branching behavior. \excludePart{The temporal profile also seems impacted.} For a 150\,ps pulsed laser beam, it is found that the E-FISH measurements are not affected by the branching. 

Non-invasive E-FISH measurements can be achieved by using (either) a lower laser power at the plasma-laser intersection and/or a shorter pulse duration. The decrease in branching intensity for a less focused laser beam (and thus lower laser power), results in an increasing accuracy of the electric profile, nonetheless, at the expense of spatial resolution. Similarly, decreasing the laser input power results in more accurate measurements with loss of signal-to-noise. However, recent work has shown that low laser power E-FISH measurements are possible~\cite{Kuhfeld_2021}.

\clearpage

\appendix
\section*{Appendix}
\FloatBarrier
\excludePart{\chapter{Deconvolution of surface charge distributions}
\label{sec: deconvolution surface charge}
\markboth{Appendix}{Appendix A: Deconvolution of surface charge distributions}

Figure~\ref{fig:deconvolutionbars} shows the application of the deconvolution model to a potential distribution on top of a grounded 1\,mm thick BSO crystal. Comsol is used to obtain the potential distribution from the surface charge distribution. The reconstructed surface charge distribution obtained by the deconvolution model described in~\cite{Mentink2022} is compared to applying the uniform field approximation, described in equation~\ref{eq:uniform}, to the potential distribution. 

\begin{figure}
    \centering
     \begin{subfigure}[b]{0.49\linewidth}
        \includegraphics[width=\textwidth]{Figures/Pockels/BarsChargeDens.png}
        \caption{}
        \label{fig:deconvolution_a}
    \end{subfigure}
    \hfill
    \begin{subfigure}[b]{0.49\linewidth}
        \includegraphics[width=\textwidth]{Figures/Pockels/BarsPotentialDist.png}
        \caption{}
        \label{fig:deconvolution_b}
    \end{subfigure}
    \vfill
    \begin{subfigure}[b]{0.49\linewidth}
        \includegraphics[width=\textwidth]{Figures/Pockels/BarsChargeRecon.png}
        \caption{}
        \label{fig:deconvolution_c}
    \end{subfigure}
    \hfill
    \begin{subfigure}[b]{0.49\textwidth}
        \includegraphics[width=\textwidth]{Figures/Pockels/BarsChargeReconUFA.png}
        \caption{ }
        \label{fig:deconvolution_d}
    \end{subfigure}
    \caption{Surface charge distribution on top of a 1\,mm BSO (a) with the corresponding  potential distribution over the BSO (b). the reconstructed charge distribution, using the the deconvolution method (c) and uniform field approximation (d). The other side of the BSO is grounded and no noise is superimposed on the potential distribution.}
    \label{fig:deconvolutionbars}
\end{figure}

\chapter{Derivation of the ISPD model equations}  \label{sec: ISPD equations}
\markboth{Appendix}{Appendix B: Derivation of the ISPD model equations}

The isothermal surface potential decay (ISPD) model is based on the isothermal relaxation current theory (IRC)~\cite{PhysRevB.7.3706} and surface potential decay measurements. According to the IRC theory, electrons are emitted from the upper half of the band gap (above the equilibrium Fermi level) to the conduction band, while hole traps are located in the lower half of the energy band (below the Fermi level) and the depth of the hole trap is referred to the valance band. This is visualised in figure~\ref{fig:trapping}.
\begin{figure}[H]
    \centering
    \includegraphics[width=1\textwidth]{Figures/Ionizer/trapping.jpg}
    \caption[]{The charge trapping and detrapping process with $E_c$ the bottom energy level of the conduction band, $E_v$ the top level of valence band and $E_F$ the Fermi level of an insulating material. Image taken from~\cite{ZHANG20061995}.}
    \label{fig:trapping}
\end{figure}

The emission rate of electrons contained in a trap level within an incremental energy range $dE$ to the conduction band is given by~\cite{PhysRevB.7.3706}
\begin{equation}
\label{eq:electronemissionperE}
\Delta {n^{'}_t} = f_0 N(E_t) e_n  \exp({-e_n t)} dE.
\end{equation}
Here, $f_0$ is the initial occupancy of traps by electrons and $N(E_t)$ the trap density at trapping energy $E_t$. The probability that a trapped electron with an energy level $E$ is emitted into the conduction band as function of time $t$ is described as
\begin{equation}
\label{eq:probabilitytrap}
e_n= \gamma \exp{(\frac{E-E_c}{kT})},
\end{equation}
where $E_c$ is the energy of the conduction band edge.
Assuming that the attempt to escape frequency $\gamma$ is in the order of $10^{11}-10^{13}$\,s$^{-1}$ and the simplification that charges in shallower traps start to be detrapped earlier than those in deeper traps~\cite{PhysRevB.7.3706}, the rate of electron emission to the conduction band is~\cite{10.1063/1.4792491}:
\begin{equation}
\label{eq:electronemission}
{n^{'}_t} = f_0 N(E_t) k T/t,
\end{equation}
with $k$ the Boltzmann constant and $T$ the temperature of the sample.

In general, the density of trap states near the surface is much higher than in the bulk of a material. If we assume that surface trap states are uniformly distributed in a thin layer, no charge is present in the bulk of the material, and if $\delta \ll L$ with $\delta$ the thickness of the top charge layer and $L$ the sample thickness, the surface potential is given by 
\begin{equation}
\label{eq:surfacepotential}
V_s = \frac{L \delta \rho}{\epsilon_0 \epsilon_r},
\end{equation}
where $\rho$ is the charge density in the top layer of the surface, $\epsilon_0$ the vacuum permittivity, and $\epsilon_r$ the relative permittivity. 
From equation~\eqref{eq:surfacepotential}, it can be observed that the charges present at the top layer of the surface determine the surface potential. As a result, mostly the surface trap states of the material are revealed by the ISPD model. The gradient of the charge decay equals~\cite{10.1063/1.4792491}
\begin{equation}
\label{eq:surfacepotentialdecay}
\frac{dV_s}{dt} = \frac{L \delta}{\epsilon_0 \epsilon_r} \frac{d\rho}{dt} = \frac{L \delta}{\epsilon_0 \epsilon_r} q {n^{'}_t},
\end{equation}
where $q$ is the elementary charge.

The electron trap density at $E_t$ is obtained by substituting equation~\eqref{eq:electronemission} in equation~\eqref{eq:surfacepotentialdecay} and rewrite the equation~\cite{10.1063/1.4792491}:
\begin{equation}
\label{eq:electrontrapdensity_app}
N(E_t) = \frac{\epsilon_0 \epsilon_r t}{k T f_0(E_t) \delta L q}\frac{dV_s}{dt}.
\end{equation}

Thus, the behavior of surface potential decay directly reflects the release of (unipolar) surface charges, which facilitates obtaining the energy distribution of electron or hole traps.

\chapter{Additional figures invasiveness} \label{sec: laser power variation}
\markboth{Appendix}{Appendix C: Additional figures invasiveness} }

\begin{figure}
    \centering
    \includegraphics[width=0.7\textwidth]{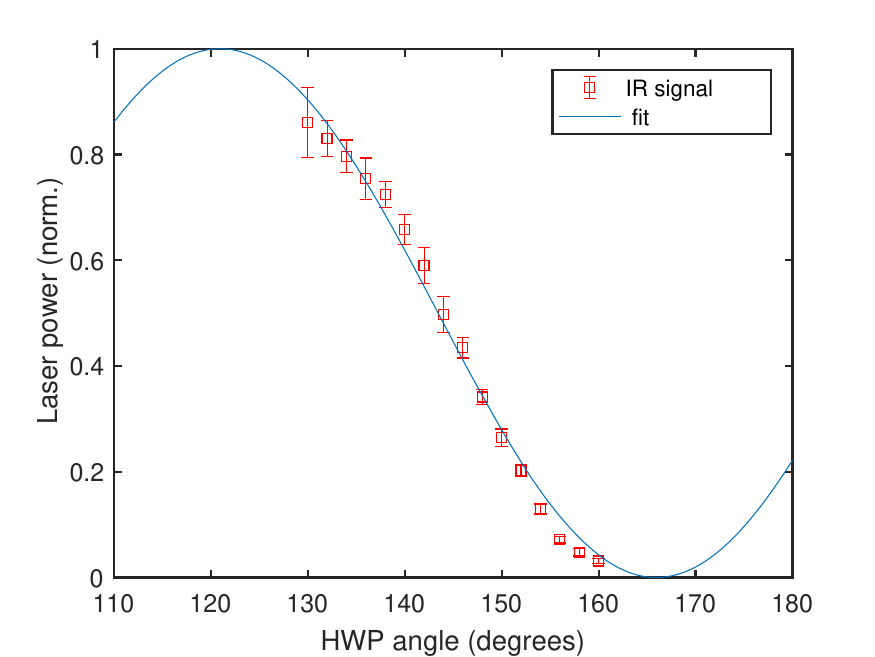}
    \caption[]{Measured IR signal (red) for varying angles of the half-wave plate for the picosecond laser, with a fit (blue) based on Malus' Law~\cite{Bennett2008} adjusted for the HWP properties. Both the data and the fit have been normalized using the maximum of the fit.}
    \label{fig:Power_fit}
\end{figure}

\excludePart{The obtained ICCD images are post-processed using Matlab by subtracting the average background noise.
A gamma correction factor $\gamma_c$ is applied to increase the visibility of the plasma~\cite{BULL201499}. The nozzle and the colors were added in post-processing for illustration purposes.
}
\begin{figure}
    \centering
    \includegraphics[width=0.7\textwidth]{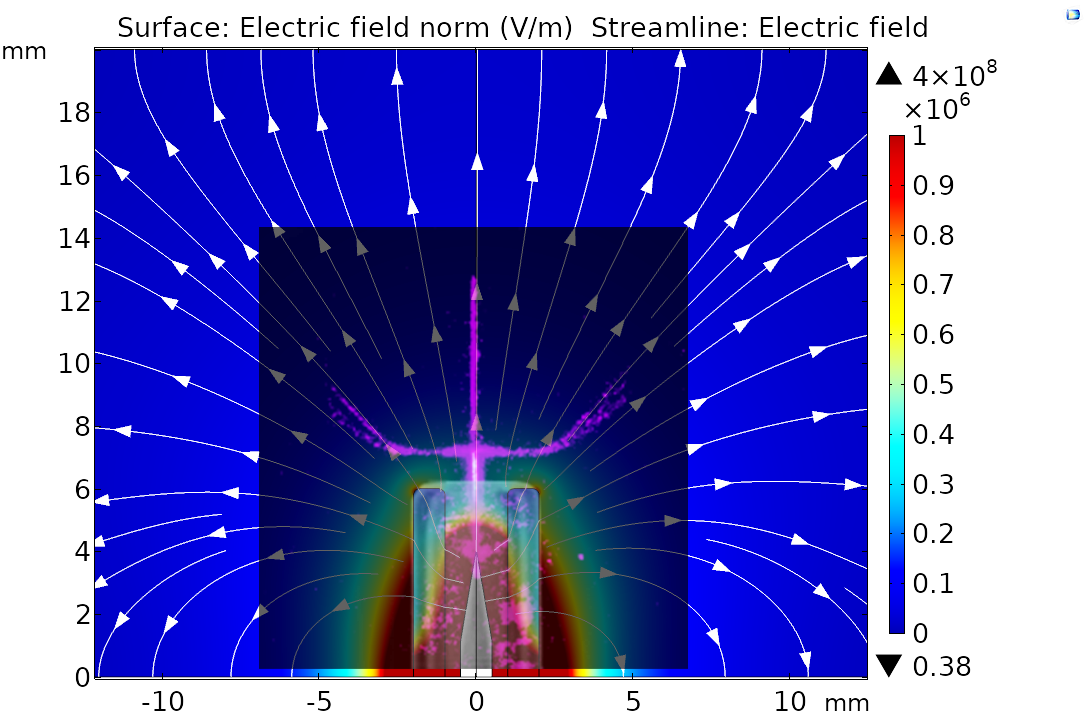}
    \caption[]{COMSOL Multiphysics model of the electric field streamlines compared to an ICCD image with gate time 220\,ns of the bullet trajectory for the picosecond laser.}
    \label{fig:streamlines_ns}
\end{figure}

\newpage

\begin{figure}
    \centering
    \includegraphics[width=1\textwidth]{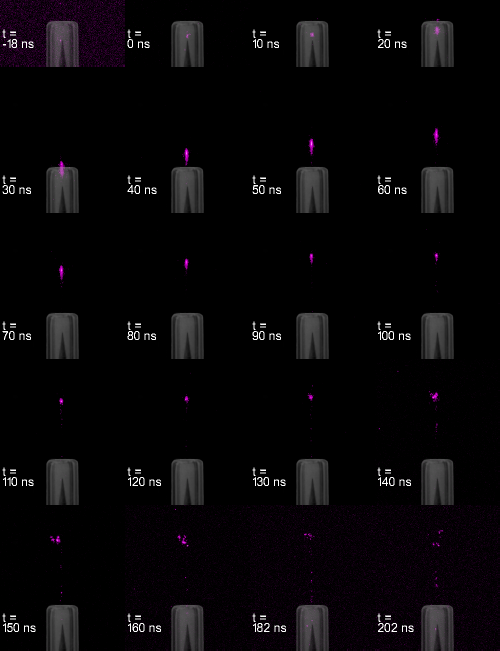}
    \caption[]{The plasma bullet trajectory with laser off, an exposure time of $2$\,ns. At $t = -18$\,ns, the HV pulse is sent to the plasma source. $\gamma_c = 1.6$.}
    \label{fig:Matrix_Loff}
\end{figure}

\begin{figure}
\centering
    \includegraphics[width=1\textwidth]{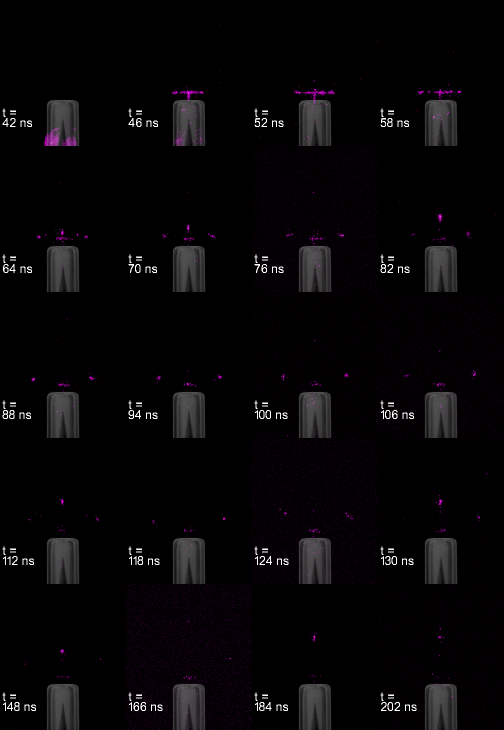}
    \caption[]{The plasma bullet with the nanosecond laser turned on. The exposure time is $2$\,ns. $\gamma_c = 1.6$.}
    \label{fig:Matrix_Jetscan_500mm_ns}
\end{figure}

\begin{figure}
\centering
    \includegraphics[width=1\textwidth]{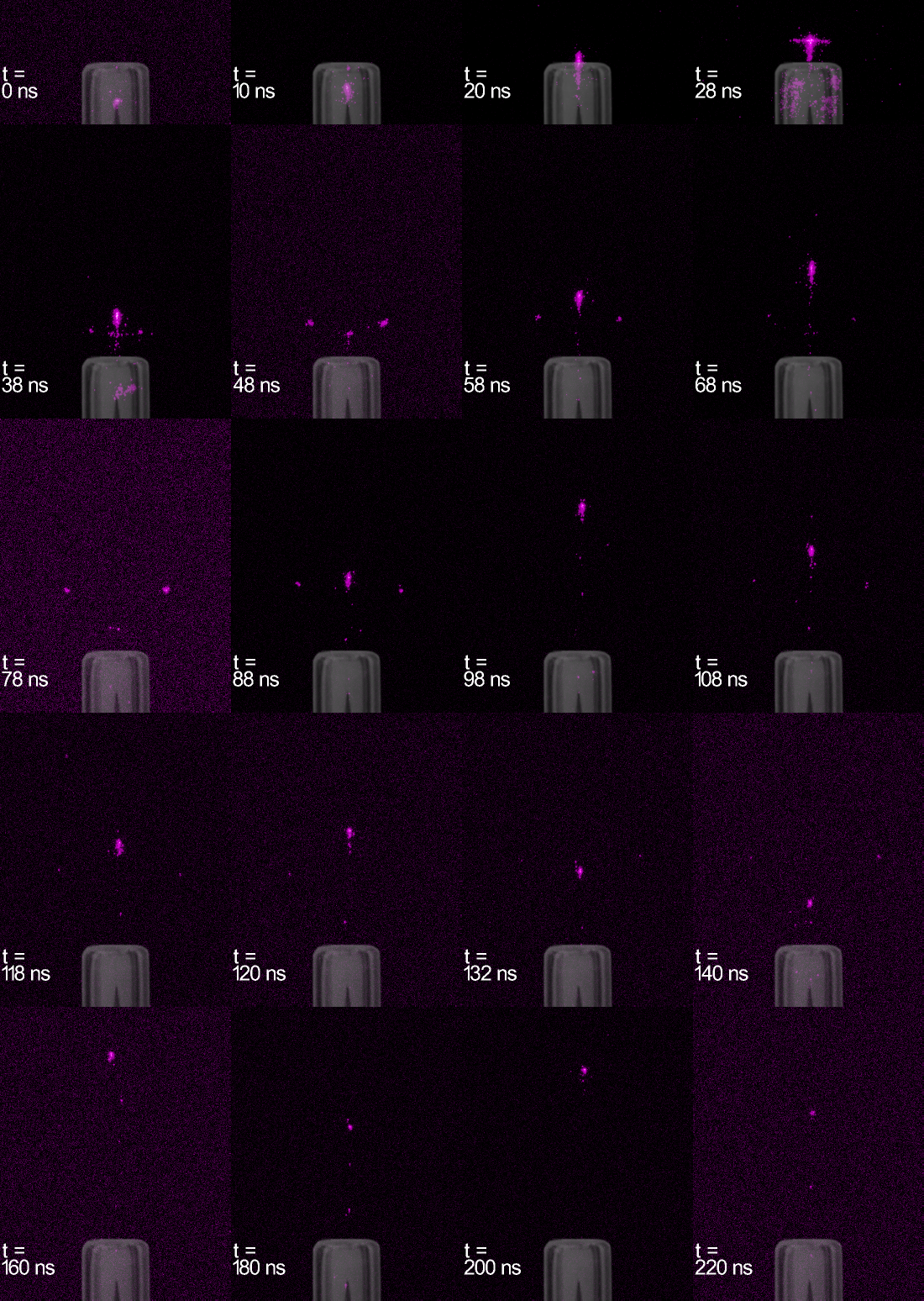}
    \caption[]{The plasma bullet with the picosecond laser turned on. The exposure time is $2$\,ns. $\gamma_c = 1.15$.}
    \label{fig:Matrix_Jetscan_500mm}
\end{figure}

\begin{figure}
\centering
    \includegraphics[width=\textwidth]{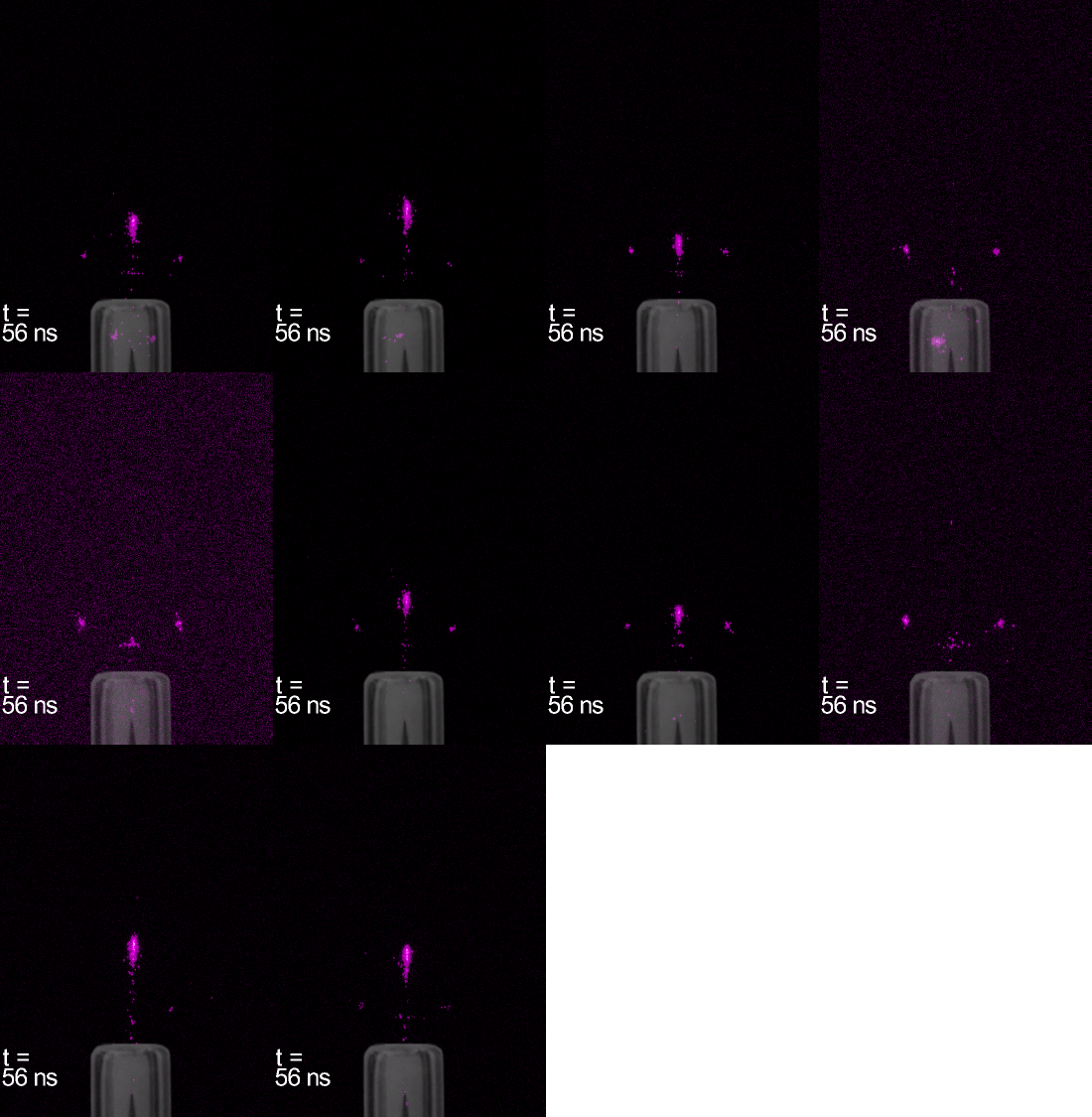}
    \caption[]{The plasma bullet at $t = 56$\,ns with the picosecond laser turned on. The exposure time is $2$\,ns. The experimental parameters do not vary, but the occurrence and intensity of the branches clearly vary between bullets.  $\gamma_c = 1.15$.}
    \label{fig:Matrix_Jetscan_stoch}
\end{figure}

\begin{figure}
\centering
    \includegraphics[width=1\textwidth]{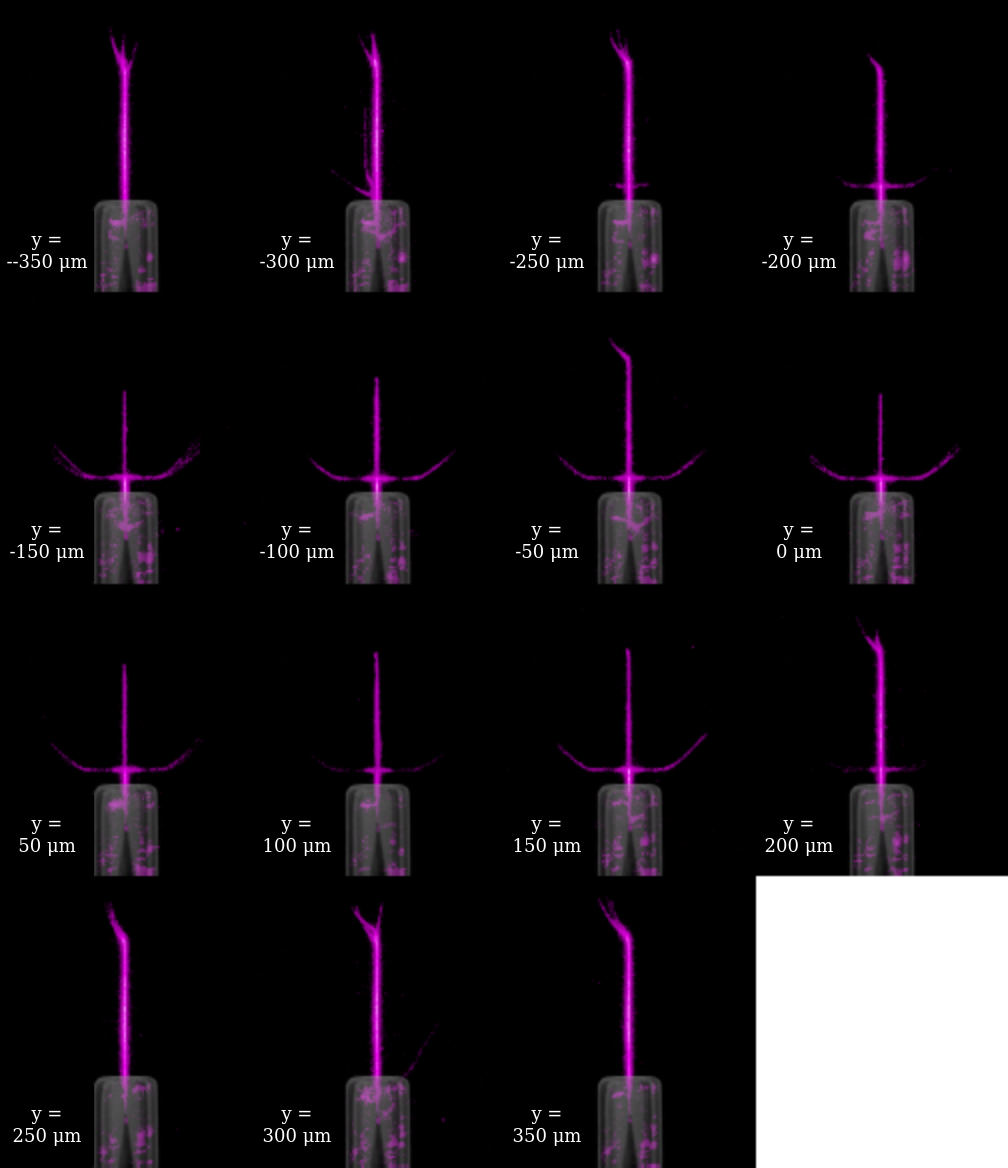}
    \caption[]{The plasma jet for varying radial positions using $170$\,ns exposure time with the nanosecond laser turned on.  $\gamma_c = 1.30$.}
    \label{fig:Matrix_hscan_horpol_500mm_ns}
\end{figure}

\begin{figure}
\centering
    \includegraphics[width=1\textwidth]{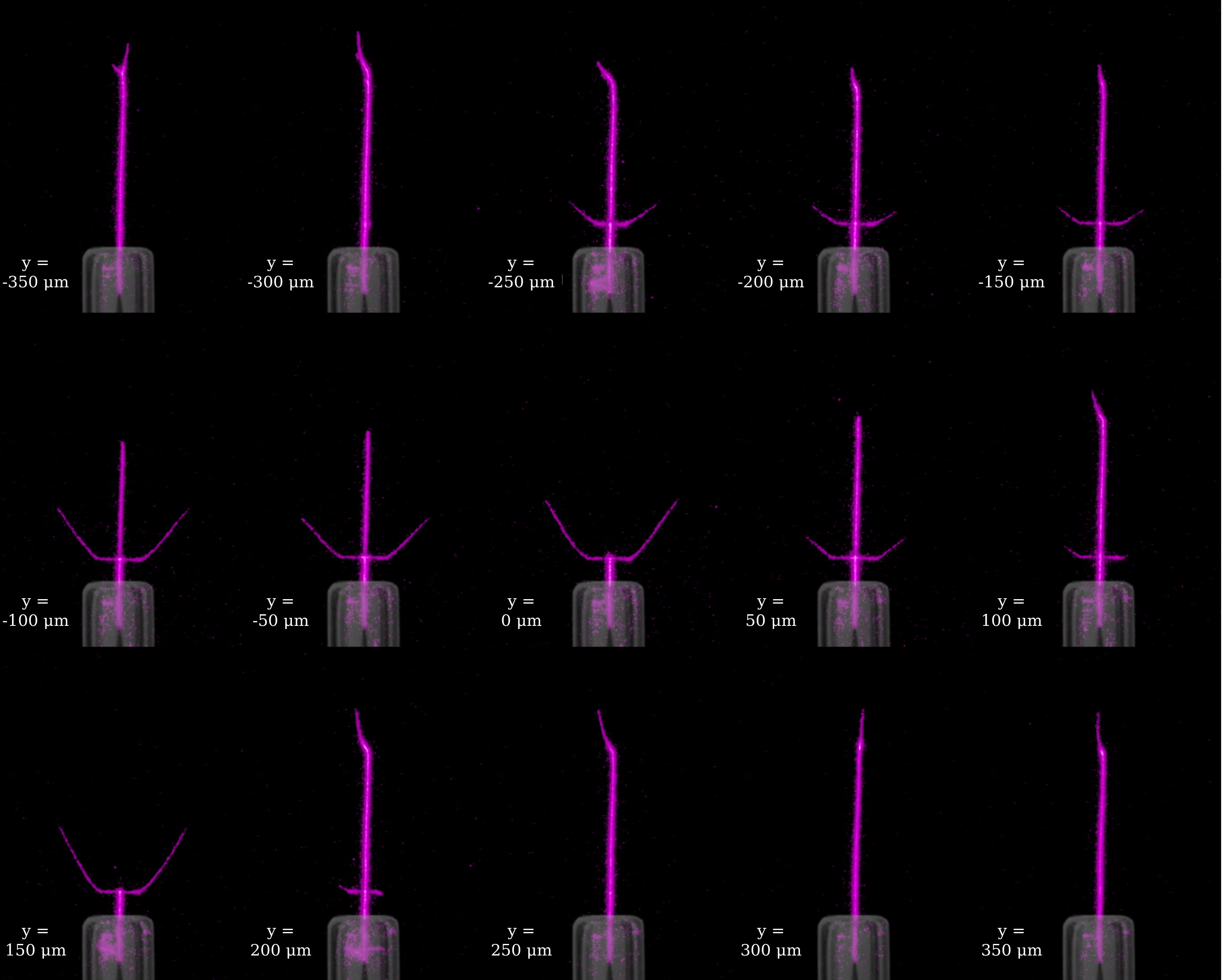}
    \caption[]{The plasma jet for varying radial positions using $220$\,ns exposure time with the picosecond laser turned on.  $\gamma_c = 1.15$.}
    \label{fig:Matrix_hscan_horpol_500mm}
\end{figure}
\excludePart{
\begin{figure}
\centering
    \includegraphics[width=0.95\textwidth]{Figures/E-FISH_invasiveness/Matrix_time_tflip_ns.png}
    \caption[]{The plasma jet for varying laser pulse delays, using $170$\,ns exposure time with the nanosescond laser turned on. $\gamma_c = 1.30$.}
    \label{fig:Matrix_time_pow1_3}
\end{figure}

\begin{figure}
\centering
    \includegraphics[width=1\textwidth]{Figures/E-FISH_invasiveness/Matrix_tscan_horpol_500mm_ps.png}
    \caption[]{The plasma jet for varying laser pulse delays, using $220$\,ns exposure time with the picosecond laser turned on. $\gamma_c = 1.15$.}
    \label{fig:Matrix_tscan_horpol_500mm}
\end{figure}}

\begin{figure}
\centering
    \includegraphics[width=1\textwidth]{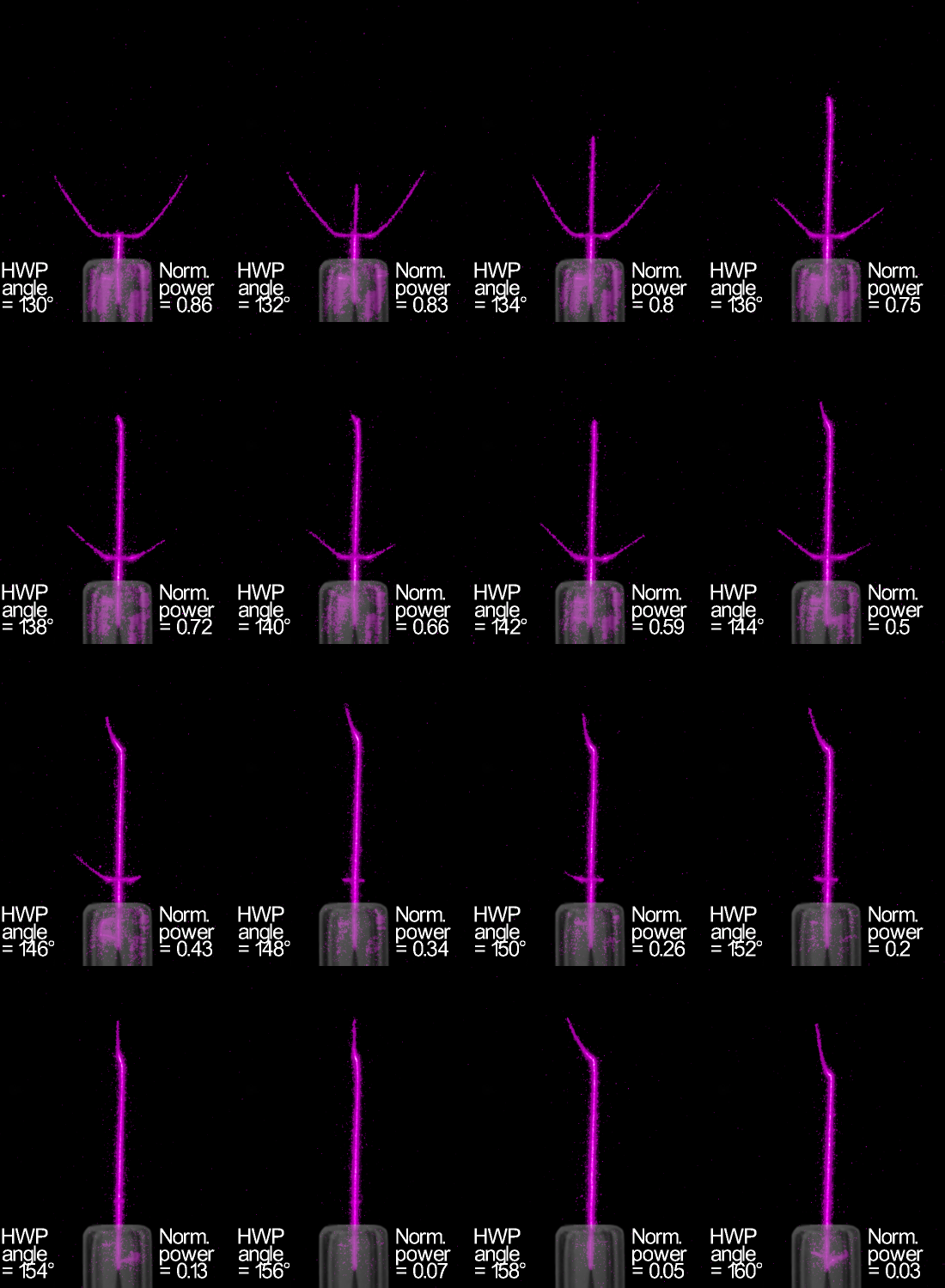}
    \caption[]{The plasma jet for varying input laser powers of the picosecond laser, using $220$\,ns exposure time. The normalized powers of the input laser for each HWP angle are given based on Figure~\ref{fig:Power_fit}. $\gamma_c = 1.15$.}
    \label{fig:Matrix_powscan_500mm}
\end{figure}

\excludePart{
\chapter{Numerical simulation model E-FISH}\label{sec:App_simulation}
\markboth{Appendix}{Appendix E: Numerical simulation model E-FISH}

For theoretical work, it is convenient to express the amplitude distribution of a Gaussian beam in a compact, albeit less intuitive form as \cite{NonlinearOpticsBook3}:

\begin{equation}\label{eq:App_GaussianBeamDistribution}
    A_n(r,z) = \frac{\mathcal{A}_n}{1 + i \zeta_n} \exp[\frac{-r^2}{w_0^2 (1 + i \zeta_n)}].
\end{equation}
Here, $r$ is the distance from the optical axis and $\mathcal{A}$ is the amplitude at the center of the focal plane, i.e. at $r = z = 0$. Moreover, $n$ indicates the frequency and $\zeta = \frac{z}{z_R}$ is a dimensionless longitudinal coordinate with $z_R$ the Rayleigh length. For E-FISH, the complex amplitude of the induced polarization is given by \cite{NonlinearOpticsBook3}:

\begin{equation}\label{eq:Theory_third_order_polarization}
    P_{2\omega}(r,z) =\frac{3}{2} \varepsilon_0 \chi^{(3)} A_{\omega}^2(r,z) E_\mathrm{ext}(z),
\end{equation}
where $\varepsilon_0$ is the vacuum permittivity and $\chi^{(3)}$ the third order electric susceptibility coefficient. 

By inserting equation~\eqref{eq:App_GaussianBeamDistribution} into equation~\eqref{eq:Theory_third_order_polarization} for a focused Gaussian beam with amplitude distribution $A_\omega$ as probing laser beam, the amplitude of the induced polarization SH wave can be determined:

\begin{equation}\label{eq:App_Polarization_GaussianBeam}
    P_{2\omega}(r,z) = \frac{3}{2} \varepsilon_0 \chi^{(3)} \frac{\mathcal{A}_\omega^2}{(1 + i \zeta_\omega)^2} \exp[\frac{-2r^2}{w_0^2 (1 + i \zeta_\omega)}] E_\mathrm{ext}(z). 
\end{equation}

In order to derive an expression of the intensity of the outgoing SH wave, the so-called driven wave equation needs to be solved first, which is given by

\begin{equation}\label{eq:App_DrivenWaveEquation}
    \bigg(2ik_{2\omega} \frac{\partial}{\partial z} + \nabla_T \bigg) A_{2\omega}(r,z) = - \frac{{(2\omega)}^2}{\varepsilon_0 c^2} P_{2\omega}(r,z) e^{-i \Delta k z},
\end{equation}
where $k_{2\omega}$ is the wavenumber of the SH wave, $\nabla_T$ is the transverse Laplacian, $A_{2\omega}$ is the amplitude distribution of the SH beam and $c$ is the speed of light in vacuum. 
To solve this equation, we will make use of the Hankel transformation, which transfers the radial space $r$ of any given function $f(r)$ into the radial wavevector space $K$. 
In this case, the Hankel tranform of zeroth-order will be applied, which is defined as \cite{HankelTransform}

\begin{equation}
    \Tilde{f}(K) = \mathcal{H}_0[f(r)] = \int_0^\infty f(r) J_0(Kr) r \mathrm{d}r,
\end{equation}
where $J_0$ is the Bessel function of the first kind of zeroth-order. 
Applying the Hankel transform to equation~\eqref{eq:App_DrivenWaveEquation} gives 

\begin{equation}
    \bigg(2 i k_{2\omega} \frac{\partial}{\partial z} - K^2 \bigg) \Tilde{A}_{2\omega}(K,z) = - \frac{(2\omega)^2}{\varepsilon_0 c^2} \Tilde{P}_{2\omega}(K,z) e^{-i \Delta k z}, 
\end{equation}
with $\Tilde{A}_{2\omega}(K,z)$ and $\Tilde{P}_{2\omega}(K,z)$ the Hankel transform of the amplitude distribution and the polarization of the induced SH wave, respectively. 
Subsequently, the left hand side of the transformed wave-equation can be written as a single term of a derivative to $z$, by multiplying both sides by $(2ik_{2\omega})^{-1} e^{iK^2z/2k_{2\omega}}$: 

\begin{equation}\label{eq:Hankel}
\begin{split}
   & \frac{\partial}{\partial z} \bigg(\Tilde{A}_{2\omega}(K,z) \exp[\frac{i K^2 z}{2 k_{2\omega}}] \bigg) = \\ &\frac{i (2\omega)^2}{2 k_{2\omega} \varepsilon_0 c^2} \exp[\frac{i K^2 z}{2 k_{2\omega}}] \Tilde{P}_{2\omega}(K,z) \exp[-i \Delta k z].
    \end{split}
\end{equation}

Next, $\Tilde{A}_{2\omega}(K,z)$ can be obtained by taking the integral of equation~\eqref{eq:Hankel} with respect to $z$ and multiplying both sides with $e^{-iK^2z/2k_{2\omega}}$:

\begin{equation}\label{eq:App_SHwaveAmplitude_Hankel}
\begin{split}
&\Tilde{A}_{2\omega}(K,z) = \\&\frac{i (2\omega)^2}{2 k_{2\omega} \varepsilon_0 c^2} \exp[-\frac{i K^2 z}{2 k_{2\omega}}] \int_{-\infty}^z \exp[\frac{i K^2 z'}{2 k_{2\omega}}] 
\Tilde{P}_{2\omega}(K,z') \exp[-i \Delta k z'] \mathrm{d}z'.
\end{split}
\end{equation}

In a later stage, the integration boundaries may be replaced by the limits to which the external electric field extends. 
Furthermore, to fully evaluate equation~\eqref{eq:App_SHwaveAmplitude_Hankel}, the Hankel transform $\Tilde{P}_{2\omega}(K,z')$ is needed. 
In this case, it can be explicitly determined from equation~\eqref{eq:App_Polarization_GaussianBeam}, resulting in: 

\begin{equation}
    \Tilde{P}_{2\omega}(K,z) = \frac{3}{8} \varepsilon_0 \chi^{(3)} \frac{\mathcal{A}_\omega^2}{1 + i \zeta_\omega} w_0^2 \exp[\frac{-w_0^2(1 + i \zeta_\omega) K^2}{8 }] E_{ext}(z).
\end{equation}

The latter two equations provide the required `tools' to (numerically) determine the amplitude distribution in the wavevector space $K$ of the outgoing SH wave induced by a given electric field $E_{ext}(z)$. 
By subsequently applying the inverse Hankel transform, the spatial properties of the SH beam can be obtained. 
However, to obtain the power contained in the SH beam, an inverse transform is not necessary. Since the intensity $I_{2\omega} (r,z)$ also depends on the
radial coordinate in this case, it is more convenient to consider the generated power $\mathcal{P}_{2\omega}(z)$, which
is obtained by integrating the intensity over its transverse cross-sectional area as
  \begin{equation}\label{eq:App_Power_SHwave_GaussianBeam}
    \mathcal{P}_{2\omega}(z) = 2\pi \int_0^\infty I_{2\omega}(r,z) r \mathrm{d}r = \pi c \varepsilon_0 n_{2\omega} \int_0^\infty |A_{2\omega}(r,z)|^2 r \mathrm{d} r,
\end{equation}
by using $I(r,z) = \frac{1}{2}c\varepsilon_0 n_{2\omega}|A(r,z)|^2$ with  $n_{2\omega}$ the refractive index for the SH wave in the medium in which the E-FISH measurement is performed.

Subsequently, by invoking Parseval's theorem  \cite{HankelTransform}, which states that  

\begin{equation}
    \int_0^\infty |A_{2\omega}(r,z)|^2 r \mathrm{d} r = \int_0^\infty |\Tilde{A}_{2\omega}(K,z)|^2 K \mathrm{d} K,
\end{equation}
the generated power of equation~\eqref{eq:App_Power_SHwave_GaussianBeam} can be written as an integral over the Hankel transform of the amplitude distribution of the generated SH wave: 

\begin{equation}\label{eq:App_final_SHpower_GaussianBeamSolution}
    \mathcal{P}_{2\omega}(z) = \pi c \varepsilon_0 n_{2\omega} \int_0^\infty 
    |\Tilde{A}_{2\omega}(K,z)|^2
    K \mathrm{d} K.
\end{equation}

This is the final expression which determines the power of the generated SH wave for the Gaussian beam solution. 
Nevertheless, equation~\eqref{eq:PWS_intensitySH} and equation~\eqref{eq:RPWS_intensitySH} 
provide the plane-wave solution and the reduced plane-wave solution in terms of intensity. 
To directly compare the PWS and RPWS to the GBS, it is convenient to rewrite the GBS in terms of intensity. 
For this purpose, equation~\eqref{eq:App_final_SHpower_GaussianBeamSolution} is divided by an effective area $a_\mathrm{eff}$, which is defined as \cite{MarcVanDerSchansPhD}

\begin{equation}\label{eq:App_EffectiveArea}
        a_\mathrm{eff} = 2\pi \int_0^\infty e^{-4r^2/w_0^2} r \mathrm{d}r = \frac{\pi w_0^2}{4}.
\end{equation}

The exponent of equation~\eqref{eq:App_EffectiveArea} represents the intensity distribution of the probing focused Gaussian beam at $z = 0$. 
The `effective intensity' of the GBS is now given by

\begin{equation}
    \label{eq:effective_intensity}I_\mathrm{2\omega}^\mathrm{GBS} =\mathcal{P}_\mathrm{2\omega}/a_\mathrm{eff}.
\end{equation}

\textsc{Matlab}'s \texttt{integral} function is used to numerically calculate the power of the GBS as given in equation~\eqref{eq:App_final_SHpower_GaussianBeamSolution}, for which the global adaptive quadrature is utilized.

The \textsc{Matlab} script is provided on the Gitlab page of the TU/e, which can be accessed via this 
\url{https://gitlab.tue.nl/s1346861/extended-analysis-on-the-influence-of-the-probing-laser-beam-and-electric-field-properties-in-e-fish-measurements}.

\chapter{Determining the Rayleigh length}\label{sec:App_rayleigh}
\markboth{Appendix}{Appendix F: Determining Rayleigh length}

To convert an E-FISH signal into an electric field value while including the effect of focusing the probing beam in the analysis, one has to know the Rayleigh length $z_R$. In this chapter, we propose a technique to obtain its value. Here, a known electric field is translated along the laser beam and, subsequently, the E-FISH signal is measured on every position along the focus, described in section~\ref{sec: focus scan} as the Gaussian beam test. The signal will depend on the wavevector mismatch and on the amount of focusing of the beam. By fitting the E-FISH signal using the numerical model with $z_R$ as a variable, the Rayleigh length can be obtained. To show the feasibility of this method, it is applied to different electric field lengths for the same beam properties ($f = 200$\,mm lens). Therefore, the Rayleigh length obtained from these different electric fields should be the same.

The distance between the position of the focus and the middle of the electrodes is called the focus displacement, $\Delta z$ in figure~\ref{fig:Methods_FocusDisplacement}. Figure~\ref{fig:Rayleigh_length} shows the normalized E-FISH signal as a function of the focus displacement for the different electric field lengths. 
Here, $\Delta k$ = 50\,m$^{-1}$, $E_0 = \frac{V}{d}$, applied voltage $V = 8\,$kV, gap distance $d = 5\,$mm, and $L = 45$\,mm, for the electric field described in equation~\eqref{eq:ExpSetup_Efield_configuration}. 
The resulting Rayleigh length determined is $3.0\pm0.8$\,mm. This corresponds to an M$^2$ factor of 1.5, which is within the specifications of the used probing laser (M$^2<2$). 
 
\begin{figure}
    \centering
    \includegraphics[width=1\textwidth]{Figures/E-FISH_analysis/ramon_zr_calibration.pdf}
    \caption{The normalized SH intensity as function of the focus displacement $\Delta z$, for different electric field lengths $L$ and a wavevector mismatch of 50\,m$^{-1}$. The experimental data is indicated with blue crosses and the fitted E-FISH signal corresponding to a Rayleigh length $z_R$ is shown in red.}
    \label{fig:Rayleigh_length}
\end{figure}

\chapter{Obtaining the weight matrix} \label{sec: weight matrix}
\markboth{Appendix}{Appendix F: Obtaining the weight matrix}

Below, we explain how we obtain the weight matrix $\mathbf{W}$ in equation~\eqref{eq:discrete-system}.
The first step is to convert the line integrals into integrals over the radial coordinate $r$, as is also done in a standard Abel transform.
Suppose we want to evaluate the following integral:
\begin{equation}
  s(x) = \int_{-\infty}^{\infty} f(\sqrt{x^2 + z^2}) g(z) \mathrm{d}z,
\end{equation}
where $r = \sqrt{x^2 + z^2}$.
We can change the integration variable to $r$, noting that
\begin{equation}
  \label{eq:var-change}
  z = \pm \sqrt{r^2 - x^2}, \quad \mathrm{d}z = \pm \frac{r \mathrm{d}r}{\sqrt{r^2 - x^2}}.
\end{equation}
Now split the integral in two parts
\begin{equation}
  s(x) = \int_{-\infty}^{0} f(\sqrt{x^2 + z^2}) g(z) \mathrm{d}z + \int_{0}^{\infty} f(\sqrt{x^2 + z^2}) g(z) \mathrm{d}z,
\end{equation}
and then change variable in both parts, using the appropriate signs from equation~\eqref{eq:var-change}:
\begin{equation}
\begin{split}
  s(x) &= \int_{\infty}^{|x|} f(r) g(-\sqrt{r^2 - x^2}) \frac{-r}{\sqrt{r^2 - x^2}}\mathrm{d}r \\
   &+ \int_{|x|}^{\infty} f(r) g(\sqrt{r^2 - x^2}) \frac{r}{\sqrt{r^2 - x^2}} \mathrm{d}r.
  \end{split}
\end{equation}
This can be simplified to
\begin{equation}
  \label{eq:transform-full}
  s(x) = \int_{|x|}^{\infty} f(r) \left[g(\sqrt{r^2 - x^2}) + g(-\sqrt{r^2 - x^2})\right] \frac{r}{\sqrt{r^2 - x^2}} \mathrm{d}r.
\end{equation}
To numerically evaluate equation~\eqref{eq:transform-full}, we for simplicity assume that the $\mathbf{f} = (f_1, f_2, \dots, f_N)$ are equally spaced, so that $r_{i+1} - r_i = \Delta r$.
We furthermore assume that $f(r)$ and $g(z)$ do not vary strongly in each interval from $r_i - 0.5 \Delta r$ to $r_i + 0.5 \Delta r$.
On the other hand, the factor $\frac{r}{\sqrt{r^2 - x^2}}$ will vary significantly (and diverge) for $r \to |x|$.
Since
\begin{equation}
  \int \frac{r}{\sqrt{r^2 - x^2}} \mathrm{d}r = \sqrt{r^2 - x^2} + C,
\end{equation}
we can handle this by analytically integrating this factor
\begin{equation}
\begin{split}
    \int_{r_i - 0.5 \Delta r}^{r_i + 0.5 \Delta r} \frac{r}{\sqrt{r^2 - x^2}} \mathrm{d}r &= \sqrt{(r_i + 0.5 \Delta r)^2 - x^2} \\&- \sqrt{(r_i - 0.5 \Delta r)^2 - x^2},
\end{split}
\end{equation}
and then changing the integral to a sum:
\begin{equation}
\begin{split}
    s(x) &\approx \sum_{i = 1}^{N} f(r_i) \left[g(\sqrt{r_i^2 - x^2}) + g(-\sqrt{r_i^2 - x^2})\right] \\&\times \left( \sqrt{(r_i + 0.5 \Delta r)^2 - x^2} - 
  \sqrt{(r_i - 0.5 \Delta r)^2 - x^2} \right).
\end{split}
  \label{eq:transform-sum-Ey}
\end{equation}
Since the integral in equation~\eqref{eq:transform-full} starts from $r = |x|$, only terms in the sum for which $r^2 - x^2 > 0$ should contribute.
This can be achieved by replacing all terms of the form $r^2 - x^2$ by zero when they are negative.

If we consider equation~\eqref{eq:sx-Er} for $E_r$, there will be an extra factor $\sin(\theta) = x/r$ in equation~\eqref{eq:transform-full}, so that
\begin{equation}
  \label{eq:integral-x}
  s(x) = \int_{|x|}^{\infty} f(r) \left[g(\sqrt{r^2 - x^2}) + g(-\sqrt{r^2 - x^2})\right] \frac{x}{\sqrt{r^2 - x^2}} \mathrm{d}r.
\end{equation}
The weight function for each sample $f_i$ can again be integrated analytically:
\begin{equation}
  \int \frac{x}{\sqrt{r^2 - x^2}} \mathrm{d}r = x \log\left(2 \sqrt{r^2 - x^2} + 2 r \right) + C.
\end{equation}
This results in a sum
\begin{equation}
\begin{split}
    &s(x) \approx \\&\sum_{i = 1}^{N} f(r_i) \left[g(\sqrt{r_i^2 - x^2}) + g(-\sqrt{r_i^2 - x^2})\right] \, x \,
  \log\left(\frac{\sqrt{r_b^2 - x^2} + r_b}{\sqrt{r_a^2 - x^2} + r_a} \right),
\end{split}
  \label{eq:transform-sum-Ex}
\end{equation}
where $r_a = r_i - 0.5 \Delta r$ and $r_b = r_i + 0.5 \Delta r$.
Note that we used $\log(b) - \log(a) = \log(b/a)$.
Again, the integral in equation~\eqref{eq:integral-x} starts from $r = |x|$, so the sum should only contain corresponding contributions.
This means that $\sqrt{r_a^2 - x^2} + r_a \geq |x|$ and similarly for $r_b$, so that these terms can be replaced by $\max(\sqrt{r_a^2 - x^2} + r_a, |x|)$.
Since $\lim_{x \to 0} x \log(x) = 0$, the case $x = 0$ should give a zero contribution.

The right-hand sides of equations~\eqref{eq:transform-sum-Ey} and \eqref{eq:transform-sum-Ex} contain the weights $w_i$ for each of the $f(r_i)$.
Computing these weights for different sampling locations $x_j$ (i.e., where $s(x)$ is measured) results in the full weight matrix $\mathbf{W}$ used in equation~\eqref{eq:discrete-system}.}

\clearpage

\section*{References}

\bibliography{References.bib}

\end{document}